\documentclass[intlimits,twoside,a4paper]{article}
\usepackage{amsmath,amssymb}
\usepackage{graphicx}
\usepackage{wrapfig}
\usepackage{bm}
\usepackage[T2A]{fontenc}
\usepackage[cp1251]{inputenc}
\usepackage{bm}

\usepackage{threeparttable}

\usepackage[eqsecnum]{cmpj}

\issue{2011}{14}{4}{43001}

\doinumber{10.5488/CMP.14.43001}

\title[Viscoelastic description of electron subsystem]%
{Viscoelastic description of electron subsystem \\of a semi-bounded
metal within generalized \\``jellium'' model}
\author[P.P. Kostrobij \textsl{et al.}]
{P.P.~Kostrobij\refaddr{label1}, B.M.~Markovych\refaddr{label1}, A.I.~Vasylenko\refaddr{label2},
M.V.~Tokarchuk\refaddr{label1,label2}}
\addresses{
\addr{label1} Lviv Polytechnic National University, 12 Bandera Str.,  79013 Lviv,  Ukraine
\addr{label2} Institute for Condensed Matter Physics of the National Academy of Sciences of Ukraine,\\
1 Svientsitskii Str., 79011 Lviv, Ukraine
}
\date{Received May 22, 2011, in final form October 13, 2011}%

\authorcopyright{P.P. Kostrobij, B.M. Markovych, A.I. Vasylenko, M.V. Tokarchuk, 2011}

\begin{document}

\maketitle


\begin{abstract}
Viscoelastic description of the electron subsystem of a
semi-bounded metal on the basis of the generalized ``jellium''
model using the method of nonequilibrium statistical Zubarev
operator is proposed. The nonequilibrium statistical operator and
the quasi-equilibrium partition function calculated by means of
the functional integration method are obtained. Transport
equations for nonequilibrium mean values of electron density
and momentum are received in the Gaussian approximation and in the
following higher approximation that corresponds to the third-order
cumulant averages in calculation of the quasi-equilibrium
partition function.
\pacs  05.60.Gg, 05.70.Np, 63.10.+a, 68.43, 82.20.Xr
\end{abstract}

\newcommand{\dd}{\mathrm{d}}
\newcommand{\ee}{{\,\textrm{e}}}
\newcommand{\ii}{\mathrm{i}}
\newcommand{\const}{\mathrm{const}}
\newcommand{\spc}{\!\!\!}
\newcommand{\Sp}{\mathop{\mathrm{Sp}}\nolimits}
\newcommand{\dint}{\displaystyle\int}
\newcommand{\zar}{\mathcal{Z}}
\newcommand{\jell}{\mathrm{jell}}
\newcommand{\ion}{\mathrm{ion}}
\newcommand{\unif}{\mathrm{unif}}
\newcommand{\ind}{\mathrm{ind}}

\section{Introduction}

Today, the studies of diffusion mechanism and catalytic reactions of
adsorbed atoms, formation of nanostructures on a metal surface are
quite topical in surface physics~\cite{l1,l11,l13,l14,l15,l16,l17,Kost000, Ign000}. In the
processes of adsorption, desorption, and surface diffusion, a
metal surface undergoes a reconstruction accompanied by a
variation of nonequilibrium properties of both electron and ion
subsystems. In this case, the electro-diffusive, viscothermal, and
electromagnetic properties of the electron subsystem change in the
field of metal surface ions. Studies of transport processes and
their particularities in the electron subsystem of semi-bounded
metals are of great importance for nanotechnologies and catalytic
technologies. Generally, the electron subsystem stays in the
states far from equilibrium during transport processes. This fact
significantly complicates the electron subsystem description. To
discover transport processes, various theoretical approaches are
developed for the spatially inhomogeneous electrons-atoms systems.
Particularly, the time-dependent density functional theory (TDDFT)
is widely used~\cite{TDF0,TDF1,TDF2,TDF3,TDF6,TDF7,TDF8,
TDF11,TDF12,TDF14,TDF15,TDF18,TDF20}. Another theoretical approach
is related to the hydrodynamic model of surface plasmons for a
spatially inhomogeneous electron gas proposed in
\cite{EQ1,EQ2,EQ3} with the use of the response theory~\cite{Grif}
based on the Boltzmann kinetic equation. The quantum statistical
theory for the description of nonequilibrium processes in the
``metal--adsorbate--gas'' systems was developed in the works~\cite{Kost5,Kost111,Kost000} using the Zubarev method of
nonequilibrium statistical operator (NSO)~\cite{l121,l122}. In
particular, a self-consistent description of nonequilibrium
processes in the atomic and the electron subsystems was presented
in~\cite{Kost5} at the kinetic level of the description of
electron processes.  To study the ionic and electron structures of
a semi-bounded metal, a generalized approach that takes the effect
of discreteness of the ion subsystem into account and is based
upon the model of a semi-bounded ``jellium''~\cite{l184,l185} was
proposed in~\cite{Kost111,Kost000}. It is worth noting that the
effect of discreteness of ionic density on the
characteristics of a semi-bounded ``jellium'' was considered in~\cite{sss1,sss2,sss3} by means of constructing a perturbation
theory with respect to the electron-ion interaction
pseudopotential. However, the linear response of the electron
subsystem to the lattice potential does not take into account the
effects of inhomogeneity of the electron subsystem. The approach
described in~\cite{l184,l185, Kost111} allows one to model the
formation of a surface potential and to calculate the partition
function for the generalized model in terms of the cumulant
averages of the ``jellium'' model. In~\cite{Kost111}, the
generalized ``jellium'' model is a basis for the statistical
description of electro-diffusive processes for the electron
subsystem of a semi-bounded metal with the use of the NSO method,
where the single parameter of the reduced description is the
nonequilibrium average value of the electron density operator. For
such a system, the quasi-equilibrium statistical sum was
calculated by means of the functional integration method for the
case of an electron-ion interaction pseudopotential. In principle,
it enables us to obtain expressions for the nonequilibrium
statistical operator in the Gaussian and in the higher
approximations with respect to the dynamic electron correlations.
In~\cite{Kost111}, the nonequilibrium statistical operator and the
generalized transport equation of inhomogeneous diffusion were
obtained for weakly nonequilibrium processes (linear approximation
with respect to the gradient of the electrochemical potential).
The same approximation is used to find the equation for the
``density-density'' time correlation function that determines the
dynamic structural factor of the electron subsystem of a
semi-bounded metal and to demonstrate the connection of this
electro-diffusive model in the linear approximation with the TDDFT~\cite{TDF0,TDF1,TDF2,TDF3}. The expressions of a nonequilibrium
statistical operator in the Gaussian and higher approximations
with respect to the dynamic electron correlations with the
quasi-equilibrium partition function calculated by means of the
functional integration method were obtained in~\cite{KMVT2011}. It
enables us to go beyond the linear approximation with respect to the
gradient of electrochemical potential. The generalized transport
equations for a nonequilibrium average of electron density
operator for strongly nonequilibrium processes in an electron
subsystem of a semi-bounded metal were presented for the
corresponding approximations of the nonequilibrium statistical
operator.

In this paper we perform a viscoelastic description of the electron
subsystem of a semi-bounded metal on the basis of the generalized
``jellium'' model, when the nonequilibrium mean values of the
electron density operator and the electron momentum density
operator are chosen for reduced description parameters. These
parameters also play an important role in
TDDFT~\cite{TDF18,TDF7}. In the second section we find the
nonequilibrium statistical operator of a viscoelastic model for
an electron subsystem of a semi-bounded metal. The general
calculation of a quasi-equilibrium partition function for a
quasi-equilibrium statistical operator of an electron subsystem
of a semi-bounded metal is obtained using a functional
integration method. The found quasi-equilibrium statistical
operator of the electron subsystem is used for the construction of
a nonequilibrium statistical operator of the system by means of the
Zubarev method~\cite{l121,l122}. In the third section we work with
the Gaussian approximation for a quasi-equilibrium partition
function, when the operators of electron density and electron
momentum density do not correlate as pair correlation functions.
Moreover, we receive a nonequilibrium statistical operator and the
corresponding equations of generalized dynamics of the
viscoelastic model for nonequilibrium averages of the
operators of electron density and their momentum density in the
Gaussian approximation. In the section 4 we use the following
higher approximation for a quasi-equilibrium partition function,
when static correlations between the operators of electron
density and electron momentum density occur with the third-order
cumulant averages. In this approximation, the nonequilibrium
statistical operator and the non-linear dynamics equations of the
viscoelastic model of an electron subsystem of a semi-bounded
metal is obtained.

\section{Nonequilibrium statistical operator for the electron
subsystem of a semi-bounded metal within the viscoelastic model}

We start with the generalized ``jellium'' model of a semi-bounded
metal that considers the effect of the ionic subsystem
discreteness. The Hamiltonian of the system could be written as follows:
\begin{eqnarray*}
  H &=&\sum\limits_{\mathbf{p},\alpha}E_\alpha(\mathbf{p})a^\dagger_\alpha(\mathbf{p}) a^{\vphantom{\dagger}}_\alpha(\mathbf{p})
       +\frac1{2SL}{\sum_{\mathbf{q}}}'\sum_k \nu_k(\mathbf{q})\rho_k(\mathbf{q})\rho_{-k}(-\mathbf{q})\nonumber
\end{eqnarray*}
\begin{eqnarray}
\label{1}
&&{} -\frac{\zar N_\ion}{SL}\sum_{\mathbf{q}}\sum_k \nu_k(\mathbf{q})S_k(\mathbf{q})\rho_{k}(\mathbf{q})
       +\frac{e N_\ion}{SL}\sum_{\mathbf{q}}\sum_k S_k(\mathbf{q})f_k(\mathbf{q})\rho_{k}(\mathbf{q})        \nonumber\\
&&{}-\frac{N}{2S}{\sum_{\mathbf{q}}}'\nu(\mathbf{q}|0)
      +\frac12\sum_{i\neq j=1}^{N_\ion}\frac1S{\sum_{\mathbf{q}}}'\zar^2\nu(\mathbf{q}|Z_i-Z_j)
      \ee^{\ii\mathbf{q}(\mathbf{R}_{\|i}-\mathbf{R}_{\|j})},
\end{eqnarray}
where $-e$ is the charge of an electron, $m$ stands for the
electron mass, $\mathbf{r}_i$, $i=1,2,\ldots,N$ is the electron
coordinate, $N_\ion$ means the number of ions in a metal with relevant
charges $\zar e$ and coordinates $\mathbf{R}_j$;
($-\infty<X_j,Y_j<+\infty$, $Z_j\leqslant Z_0$, $Z_0=\const$), $z=Z_0$
is a dividing surface, $V=SL$ is a volume of the system, $S$ means
a surface square of a semi-bounded metal, $L$ denotes an area of a
normal coordinate changing: $z\in(-L/2,+L/2)$, $S\to\infty$,
$L\to\infty$. $E_\alpha(\mathbf{p})=\frac{\hbar^2p^2}{2m}+\varepsilon_\alpha$ is
the electron energy in the state $(\mathbf{p},\alpha)$. The
hachure at sum denotes the absence of terms with $\mathbf{q}=0$
(two-dimension wavevector in a semi-bounded metal surface) due to the electro-neutrality condition $\zar N_\ion=N$.
$\nu_k(\mathbf{q})=4\pi e^2/(q^2+k^2)$ is the Fourier-image of
interaction potential between electrons, $f_k(\mathbf{q})$ is the
Fourier-image of the model pseudopotential of interaction between
ions and electrons $w(\mathbf{r}_i-\mathbf{R}_j)=-\frac{\zar
e}{|\mathbf{r}_i-\mathbf{R}_j|}+f(\mathbf{r}_i-\mathbf{R}_j)$,
$\mathbf{R}_{\|j}=(X_j,Y_j)$, $\nu(\mathbf{q}|z)=2\pi
e^2\ee^{-q|z|}/q$ means the two-dimension Fourier-image of the
Coulomb potential, \begin{equation}\label{2}
S_k(\mathbf{q})=\frac1{N_\ion}\sum\limits_{j=1}^{N_\ion}\ee^{-\ii\mathbf{q}\mathbf{R}_{\|j}-\ii
k Z_j}
\end{equation} is the structure factor of the ionic subsystem, the
Fourier-components of electrons density, the ``collective''
variable is
\begin{equation}\label{3}
\rho_k(\mathbf{q})=\sum_{\mathbf{p},\alpha_1,\alpha_2}
\langle\alpha_1|\ee^{\ii k z}|\alpha_2\rangle
a^\dagger_{\alpha_1}(\mathbf{p})
a^{\vphantom{\dagger}}_{\alpha_2}(\mathbf{p}-\mathbf{q}),
\end{equation}
where \[ \langle\alpha_1|\ldots|\alpha_2\rangle=\int\!\!\dd z\,
\varphi^*_{\alpha_1}(z)\ldots\varphi^{\vphantom{*}}_{\alpha_2}(z),
 \]
$\varphi_\alpha(z)$ and $\varepsilon_\alpha$ are the
eigenfunctions and the eigenvalues of the Schrodinger equation
\[
  \left[
   -\frac{\hbar^2}{2m}\frac{\dd^2}{\dd z^2}+V(z)
  \right]\varphi_\alpha(z)=\varepsilon_\alpha\varphi_\alpha(z),
 \]
$V(\mathbf{r})=V(z)$ is a surface potential that is a function of
only normal electron coordinate.

The averages of the electrons density operator
$\langle\rho(\mathbf{r})\rangle^t$ and their momentum density
$\langle\mathbf{p}(\mathbf{r})\rangle^t$ could be chosen for the
main parameters of the reduced description for the study of
viscoelastic processes within the formulated model. These
parameters are connected with the relevant inhomogeneous electric
and magnetic fields:
 \begin{equation}\label{4}
    \begin{split}
      {\bm{\nabla}}\cdot\langle\mathbf{E}(\mathbf{r})\rangle^t & =e\langle\rho(\mathbf{r})\rangle^t, \\
      {\bm{\nabla}}\times\langle\mathbf{H}(\mathbf{r})\rangle^t & =-\frac1c\frac{\partial}{\partial t}
       \langle\mathbf{E}(\mathbf{r})\rangle^t
      +\frac{4\pi}c\frac{e}{m}\langle\mathbf{p}(\mathbf{r})\rangle^t,
    \end{split}
 \end{equation}
where $c$ is the speed of light,
$\langle\ldots\rangle^t=\Sp[\ldots\rho(t)]$, $\rho(t)$ means the
nonequilibrium statistical operator of the generalized ``jellium''
model, which satisfies the Liouville equation with the Hamiltonian
(\ref{1}). The solution of the Liouville equation for $\rho(t)$ in
the Zubarev method with taking into account a projecting technique
can be presented in a general form:
 \begin{equation}\label{5}
    \rho(t)=\rho_{\mathrm{q}}(t)-\int\limits_{-\infty}^t\!\!\!\!\ee^{\varepsilon(t-t')}
    T_{\mathrm{q}}(t,t')\big[1-\mathcal{P}_{\mathrm{q}}(t')\big]\ii L_N\rho_{\mathrm{q}}(t')\dd t',
 \end{equation}
where
 \[T_{\mathrm{q}}(t,t')=\exp_+\left\{-\int\limits_{t'}^t\big[1-\mathcal{P}_{\mathrm{q}}(t'')\big]\ii L_N\dd t''\right\}\]
denotes the generalized evolution operator with taking into
account a projecting technique, $\mathcal{P}_{\mathrm{q}}(t')$ is the
generalized Kawasaki-Gunton projection operator. Its structure
depends on the quasi-equilibrium statistical operator $\rho_{\mathrm{q}}(t)$.
In our case $\mathcal{P}_{\mathrm{q}}(t)$ could be written down as:
 \begin{eqnarray}\label{6}
     \mathcal{P}_{\mathrm{q}}(t)\rho' & =&\Bigg(\rho_{\mathrm{q}}(t)-\sum_{\mathbf{q},k}\frac{\delta \rho_{\mathrm{q}}(t)}{\delta\langle\rho_k(\mathbf{q})\rangle^t}
                              \langle\rho_k(\mathbf{q})\rangle^t
                              -\sum_{\mathbf{q},k}\frac{\delta
                              \rho_{\mathrm{q}}(t)}{\delta\langle\mathbf{p}_k(\mathbf{q})\rangle^t}\cdot
                              \langle\mathbf{p}_k(\mathbf{q})\rangle^t\Bigg)\Sp\rho'  \nonumber\\
       & + & \sum_{\mathbf{q},k}\frac{\delta \rho_{\mathrm{q}}(t)}{\delta\langle\rho_k(\mathbf{q})\rangle^t}
                              \Sp\rho_k(\mathbf{q})\rho'
         +\sum_{\mathbf{q},k}\frac{\delta
         \rho_{\mathrm{q}}(t)}{\delta\langle\mathbf{p}_k(\mathbf{q})\rangle^t}\cdot
                              \Sp\mathbf{p}_k(\mathbf{q})\rho'
 \end{eqnarray}
and it has the following properties
 \[
  \mathcal{P}_{\mathrm{q}}(t)\rho(t)=\rho_{\mathrm{q}}(t),\qquad
  \mathcal{P}_{\mathrm{q}}(t)\rho_{\mathrm{q}}(t')=\rho_{\mathrm{q}}(t),\qquad
  \mathcal{P}_{\mathrm{q}}(t)\mathcal{P}_{\mathrm{q}}(t')=\mathcal{P}_{\mathrm{q}}(t),\qquad
  \big(1-\mathcal{P}_{\mathrm{q}}(t)\big)\mathcal{P}_{\mathrm{q}}(t)=0.
 \]
We define the quasi-equilibrium statistical operator $\rho_{\mathrm{q}}(t)$
with the principle of the Gibbs entropy maximum when the
parameters of the reduced description
$\langle\rho(\mathbf{r})\rangle^t$ and
$\langle\mathbf{p}(\mathbf{r})\rangle^t$ are fixed and the
normalization condition $\Sp\rho_{\mathrm{q}}(t)=1$ is satisfied. In our case
we find
 \begin{equation}\label{7}
    \rho_{\mathrm{q}}(t)=\exp
    \left(
     -\Phi(t)-\beta
     \left\{
      H-\frac1{SL}\sum_{\mathbf{q},k}
      \Big[
       \overline{\mu}_k(\mathbf{q};t)\rho_k(\mathbf{q};t)+\mathbf{A}_k(\mathbf{q};t)\cdot\mathbf{p}_k(\mathbf{q};t)
      \Big]
     \right\}
    \right),
 \end{equation}
 \begin{equation}\label{8}
    \Phi(t)=\ln\Sp\exp
    \left(
     -\beta
     \left\{
      H-\frac1{SL}\sum_{\mathbf{q},k}
      \Big[
       \overline{\mu}_k(\mathbf{q};t)\rho_k(\mathbf{q};t)+\mathbf{A}_k(\mathbf{q};t)\cdot\mathbf{p}_k(\mathbf{q};t)
      \Big]
     \right\}
    \right),
 \end{equation}
where
 $\overline{\mu}_k(\mathbf{q};t)={\mu}_k(\mathbf{q};t)+e\varphi_k(\mathbf{q};t)$
stands for the Fourier-component of the electrochemical potential
of electrons,
 ${\mu}_k(\mathbf{q};t)$
is the Fourier-component of the chemical potential,
 $\varphi_k(\mathbf{q};t)$ is the Fourier-component of the local electrical potential,
 $\mathbf{A}_k(\mathbf{q};t)=\mathbf{v}_k(\mathbf{q};t)-c^{-1}\mathbf{a}_k(\mathbf{q};t)$,
 $\mathbf{v}_k(\mathbf{q};t)$ denotes the Fourier-component of the average velocity of electrons,
 $\mathbf{a}_k(\mathbf{q};t)$ is the Fourier-component of the vector potential $\mathbf{a}(\mathbf{r};t)$
of electromagnetic field:
 \begin{equation}\label{9}
      \langle\mathbf{H}(\mathbf{r})\rangle^t  =\bm{\nabla}\times\mathbf{a}(\mathbf{r},t), \qquad
      \langle\mathbf{E}^t(\mathbf{r})\rangle^t  =-\frac1c\frac{\partial}{\partial t}\mathbf{a}(\mathbf{r},t),
 \end{equation}
where $\langle\mathbf{E}^t(\mathbf{r})\rangle^t$ is a tangential
part of electrical field, $\overline{\mu}_k(\mathbf{q};t)$ is
defined with the self-consistent condition:
 \begin{equation}\label{10}
    \langle\rho_k(\mathbf{q})\rangle^t=
    \langle\rho_k(\mathbf{q})\rangle^t_{\mathrm{q}}
 \end{equation}
and the thermodynamical relations:
 \begin{equation}\label{11}
    \frac{\delta\Phi(t)}{\delta\frac{\beta}{SL}\overline{\mu}_k(\mathbf{q};t)}=\langle\rho_k(\mathbf{q})\rangle^t,
 \end{equation}
 \begin{equation}\label{12}
    \frac{\delta S(t)}{\delta\langle\rho_k(\mathbf{q};t)\rangle^t}=-\frac{\beta}{SL}\mu_k(\mathbf{q};t),\qquad
    \frac{\delta S(t)}{\delta\langle e\rho_k(\mathbf{q};t)\rangle^t}=-\frac{\beta}{SL}\varphi_k(\mathbf{q};t),
 \end{equation}
when $\langle \mathbf{p}_k(\mathbf{q};t)\rangle^t$ are fixed,
where
 $S(t)$ is the nonequilibrium entropy defined using the Gibbs method:
 \begin{eqnarray}\label{13}
    S(t) &=&-\Sp\big[\ln\rho_{\mathrm{q}}(t)\big]\rho_{\mathrm{q}}(t) \nonumber\\
         &=&\Phi(t)+\beta
         \left\{
           \langle H\rangle^t
           -\frac1{SL}\sum_{\mathbf{q}}\sum_k
           \Big[
            \overline{\mu}_k(\mathbf{q};t)\langle\rho_k(\mathbf{q})\rangle^t_{\mathrm{q}}
            +
            \mathbf{A}_k(\mathbf{q};t)\cdot\langle \mathbf{p}_k(\mathbf{q})\rangle^t_{\mathrm{q}}
           \Big]
         \right\}\nonumber \\
         &=&\ln Z(t)+\beta\left\{
           \langle H\rangle^t
           -\frac1{SL}\sum_{\mathbf{q}}\sum_k
           \Big[
            \overline{\mu}_k(\mathbf{q};t)\langle\rho_k(\mathbf{q})\rangle^t
            +
            \mathbf{A}_k(\mathbf{q};t)\cdot\langle \mathbf{p}_k(\mathbf{q})\rangle^t
           \Big]
         \right\},
 \end{eqnarray}
and besides
 \begin{equation}\label{13a}
    \frac{\delta\Phi(t)}{\delta\frac{\beta}{SL} \mathbf{A}_k(\mathbf{q};t)}=\langle \mathbf{p}_k(\mathbf{q})\rangle^t,
 \end{equation}
 \begin{equation}\label{14}
    \frac{\delta S(t)}{\delta \langle \mathbf{p}_k(\mathbf{q})\rangle^t}=-\frac{\beta}{SL}\langle \mathbf{v}_k(\mathbf{q})\rangle^t,\qquad
    \frac{\delta S(t)}{\delta \langle\frac1c\mathbf{p}_k(\mathbf{q})\rangle^t}=- \frac{\beta}{SL}\mathbf{a}_k(\mathbf{q};t).
 \end{equation}
 $Z(t)$ is the partition function of the
quasi-equilibrium statistical operator:
 \begin{equation}\label{15}
    Z(t)=\Sp\exp
    \left(
     -\beta
     \left\{
      H-\frac1{SL}\sum_{\mathbf{q},k}
      \Big[
       \overline{\mu}_k(\mathbf{q};t)\rho_k(\mathbf{q})+\mathbf{A}_k(\mathbf{q};t)\cdot\mathbf{p}_k(\mathbf{q})
      \Big]
     \right\}
    \right).
 \end{equation}
One should calculate the quasi-equilibrium partition function
(\ref{15}) to define the structure of the Kawasaki-Gunton projection
operator, to find the nonequilibrium parameters
$\overline{\mu}_k(\mathbf{q};t)$, $\mathbf{A}_k(\mathbf{q};t)$,
and thus, to obtain the nonequilibrium statistical operator
$\rho(t)$. The approach based on the ``jellium'' model
 is proposed in~\cite{Kost111,KMVT2011} in the diffusion
description case, where the average value of the electron density
operator $\langle\rho(\mathbf{r})\rangle^t$ is chosen for the
reduced description parameter. We use this approach to calculate
$Z(t)$ (\ref{15}). We apply the functional integration method to
present $Z(t)$ as follows:
 \begin{equation}\label{16}
    Z(t)=\exp
    \left[
     \beta\frac{N}{2S}{\sum_{\mathbf{q}}}'\nu(\mathbf{q}|0)
    \right]
    Z_\jell
        \Delta Z(t).
 \end{equation}
Here,
 \begin{equation}\label{17}
    Z_\jell=\Sp
    \left[
     \exp(-\beta H_0) {\rm T} S_1(\beta)
    \right]
 \end{equation}
is the partition function of the ``jellium'' model of the electron
subsystem that corresponds to the equilibrium state, found in
\cite{Kost000,Kost111,Kost22,l184}.
 \begin{equation}\label{18}
    S_1(\beta)=\exp
    \left[
     -\frac1{S L}\int\limits_0^\beta
     \dd\beta'
      {\sum\limits_{\mathbf{q}}}'
      \sum\limits_k
      \nu_k(\mathbf{q})\rho_k(\mathbf{q}|\beta')\rho_{-k}(-\mathbf{q}|\beta')
    \right]
 \end{equation}
denotes the contribution of electron interaction and
$\rho_k(\mathbf{q}|\beta')=\ee^{\beta'H_0}\rho_k(\mathbf{q})\ee^{-\beta'H_0}$,
 \begin{equation}\label{19}
    \Delta Z(t)=\frac1{Z_\jell}
    \Sp
    \left[
     \exp(-\beta H_0) {\rm T} S_1(\beta) S_2(\beta;t)
    \right]
    = \langle S_2(\beta;t)\rangle_\jell\,,
 \end{equation}
where
  \[
   \langle (\ldots)\rangle_\jell=\frac1{Z_\jell}
    \Sp
    \left[
     \exp(-\beta H_0) {\rm T} S_1(\beta) (\ldots)
    \right],
  \]
 \begin{equation}\label{20}
    S_2(\beta;t)={\rm T}
    \exp
    \left[
     -\frac1{S L}
     \int\limits_0^\beta\dd\beta'
     \sum_{k,\mathbf{q}}
     \widetilde{B}(\mathbf{q},k;t)
     \widetilde{W}^{(+)}_k(\mathbf{q};\beta')
    \right],
 \end{equation}
and
 \[
  \widetilde{C}(\mathbf{q},k;t)=
  \mathrm{col}\left[B_k(\mathbf{q};t), \mathbf{A}_k(\mathbf{q};t)\right]
 \]
is a column vector, $B_k(\mathbf{q};t)=N_\ion
S_k(\mathbf{q})w_k(\mathbf{q})-\overline{\mu}_k(\mathbf{q};t)$,
 \[
  \widetilde{W}^{(+)}_k(\mathbf{q};\beta)=
  \left[
   \rho_k(\mathbf{q};\beta),\,\mathbf{p}_k(\mathbf{q};\beta)
  \right]
 \]
is a row vector. One can write down $\Delta Z(t)$ after applying
the cumulant representation:
 \begin{eqnarray}\label{21}
   \Delta Z(t) & =&\exp\Bigg[\sum_{n=1}^\infty\frac{\ii^n}{n!}\left(\frac{\beta}{S L}\right)^n
       \sum_{k_1,\ldots,k_n}\sum_{\mathbf{q}_1,\ldots,\mathbf{q}_n}
       \widetilde{C}(\mathbf{q}_1,k_1;t)\ldots         \nonumber\\
     &\times & \widetilde{C}(\mathbf{q}_n,k_n;t)
       \widetilde{{\mathfrak M}}_{-k_1,\ldots,-k_n}(-\mathbf{q}_1,\ldots,-\mathbf{q}_n;\beta)\Bigg],
 \end{eqnarray}
where
 \begin{equation}\label{22}
    \widetilde{{\mathfrak M}}_{k_1,\ldots,k_n}(\mathbf{q}_1,\ldots,\mathbf{q}_n;\beta)
    =\ii^n\left\langle{\rm T}\widetilde{W}_{k_1}(\mathbf{q}_1;\beta)\ldots \widetilde{W}_{k_n}(\mathbf{q}_n;\beta)\right\rangle_\jell^{\mathrm{c}}
 \end{equation}
denote matrices of fluctuations of the cumulant averages of electron
density and electron momentum, which are obtained with the
nonequilibrium statistical operator of the ``jellium'' model of
the electron subsystem~\cite{Kost000,Kost111,l184,l185}.
Particularly, the matrix of the second cumulant has the following
structure:
 \begin{equation}\label{23}
    \widetilde{{\mathfrak M}}_{k_1,k_2}(\mathbf{q}_1,\mathbf{q}_2;\beta)=
    \left(
      \begin{array}{cc}
        {\mathfrak M}_{k_1,k_2}^{\rho\rho}(\mathbf{q}_1,\mathbf{q}_2;\beta) & \bm{\mathfrak M}_{k_1,k_2}^{\rho\mathbf{p}}(\mathbf{q}_1,\mathbf{q}_2;\beta) \\
        \bm{\mathfrak M}_{k_1,k_2}^{\mathbf{p}\rho}(\mathbf{q}_1,\mathbf{q}_2;\beta) & \bm{\mathfrak M}_{k_1,k_2}^{\mathbf{p}\mathbf{p}}(\mathbf{q}_1,\mathbf{q}_2;\beta) \\
      \end{array}
    \right),
 \end{equation}
where
 \begin{equation}\label{24}
    {\mathfrak M}_{k_1,k_2}^{\rho\rho}(\mathbf{q}_1,\mathbf{q}_2;\beta)=
    \langle\rho_{k_1}(\mathbf{q}_1;\beta)\rho_{k_2}(\mathbf{q}_2;\beta)\rangle_\jell-
    \langle\rho_{k_1}(\mathbf{q}_1;\beta)\rangle_\jell
    \langle\rho_{k_2}(\mathbf{q}_2;\beta)\rangle_\jell\,,
 \end{equation}
 \begin{equation}\label{25}
    \bm{\mathfrak M}_{k_1,k_2}^{\rho\mathbf{p}}(\mathbf{q}_1,\mathbf{q}_2;\beta)=
    \langle\rho_{k_1}(\mathbf{q}_1;\beta)\mathbf{p}_{k_2}(\mathbf{q}_2;\beta)\rangle_\jell-
    \langle\rho_{k_1}(\mathbf{q}_1;\beta)\rangle_\jell
    \langle\mathbf{p}_{k_2}(\mathbf{q}_2;\beta)\rangle_\jell\,,
 \end{equation}
 \begin{equation}\label{26}
    {\bm{\mathfrak M}}_{k_1,k_2}^{\mathbf{p}\mathbf{p}}(\mathbf{q}_1,\mathbf{q}_2;\beta)=
    \langle\mathbf{p}_{k_1}(\mathbf{q}_1;\beta)\mathbf{p}_{k_2}(\mathbf{q}_2;\beta)\rangle_\jell-
    \langle\mathbf{p}_{k_1}(\mathbf{q}_1;\beta)\rangle_\jell
    \langle\mathbf{p}_{k_2}(\mathbf{q}_2;\beta)\rangle_\jell\,,
 \end{equation}
and $\langle\mathbf{p}_{k_i}(\mathbf{q}_i;\beta)\rangle_\jell=0$,
since averaging of the momentum density operator proceeds with the
equilibrium statistical operator. For the same reason, the averages
 \[
  \langle\rho_{k_1}(\mathbf{q}_1;\beta)\mathbf{p}_{k_2}(\mathbf{q}_2;\beta)\rangle_\jell=
  \langle\mathbf{p}_{k_1}(\mathbf{q}_1;\beta)\rho_{k_2}(\mathbf{q}_2;\beta)\rangle_\jell=0,
 \]
therefore, matrix (\ref{23}) is a diagonal one:
 \begin{equation}\label{27}
    \widetilde{{\mathfrak M}}_{k_1,k_2}(\mathbf{q}_1,\mathbf{q}_2;\beta)=
    \left(
      \begin{array}{cc}
        {\mathfrak M}_{k_1,k_2}^{\rho\rho}(\mathbf{q}_1,\mathbf{q}_2;\beta) & 0 \\
        0 & \bm{\mathfrak M}_{k_1,k_2}^{\mathbf{p}\mathbf{p}}(\mathbf{q}_1,\mathbf{q}_2;\beta) \\
      \end{array}
    \right).
 \end{equation}
Considering the above, one can present $\Delta Z(t)$ in the
Gaussian approximation as
 \begin{eqnarray}\label{28}
   \Delta Z^{\rm G}(t) & =&\exp\Bigg[-\frac12\left(\frac{\beta}{S L}\right)^2
       \sum_{k_1,k_2}\sum_{\mathbf{q}_1,\mathbf{q}_2}
       \bigg(B_{k_1}(\mathbf{q}_1;t)B_{k_2}(\mathbf{q}_2;t){\mathfrak M}_{k_1,k_2}^{\rho\rho}(\mathbf{q}_1,\mathbf{q}_2;\beta)         \nonumber       \\
     &+& \mathbf{A}_{k_1}(\mathbf{q}_1;t)\cdot\bm{\mathfrak M}_{k_1,k_2}^{\mathbf{p}\mathbf{p}}(\mathbf{q}_1,\mathbf{q}_2;\beta)
        \cdot\mathbf{A}_{k_2}(\mathbf{q}_2;t)\bigg)\Bigg].
 \end{eqnarray}
In the next section we find the nonequilibrium statistical
operator in the Gaussian approximation, where the operators of
electron density and electron momentum density do not correlate as
a pair correlation function.

 \section{The Gaussian approximation}

Let us present the relevant statistical operator $\rho_{\mathrm{q}}(t)$  to
find the nonequilibrium statistical operator in the Gaussian
approximation. Considering (\ref{28}) one can write down:
 \begin{equation}\label{29}
    \rho^{\mathrm{(G)}}_{\mathrm{q}}(t)=\exp\left[-\widehat{S}^{\rm G}(t)\right],
 \end{equation}
where
 \begin{eqnarray}\label{30}
      \widehat{S}^{\rm G}(t) & =&\beta\frac{N}{2S}{\sum\limits_{\mathbf{q}}}'\nu(\mathbf{q}|0)+\ln Z_\jell \nonumber\\
        & -& \frac12\left(\frac{\beta}{S L}\right)^2
        \sum\limits_{k_1,k_2}
        \sum\limits_{\mathbf{q}_1,\mathbf{q}_2}
        \Big[
         B_{k_1}(\mathbf{q}_1;t)B_{k_2}(\mathbf{q}_2;t){\mathfrak M}_{k_1,k_2}^{\rho\rho}(\mathbf{q}_1,\mathbf{q}_2) \nonumber\\
        & +& \mathbf{A}_{k_1}(\mathbf{q}_1;t)\cdot\bm{\mathfrak M}_{k_1,k_2}^{\mathbf{p}\mathbf{p}}(\mathbf{q}_1,\mathbf{q}_2)\cdot
        \mathbf{A}_{k_2}(\mathbf{q}_2;t)\Big] \nonumber\\
        & +& \beta \bigg\{H-\frac1{S L}\sum\limits_{k,\mathbf{q}}
          \Big[\overline{\mu}_k(\mathbf{q};t)\rho_k(\mathbf{q})+\mathbf{A}_k(\mathbf{q};t)\cdot\mathbf{p}_k(\mathbf{q})\Big]
         \bigg\}
 \end{eqnarray}
is the entropy operator. In order to exclude the parameters
$\overline{\mu}_k(\mathbf{q};t)$ we use the following
thermodynamical relation:
 \[
  \frac{\delta \Phi^{\mathrm{(G)}}(t)}{\delta \frac{\beta}{SL}\overline{\mu}_k(\mathbf{q};t)}=
  \langle\rho_k(\mathbf{q})\rangle^t,
 \]
from which one can find:
 \begin{equation}\label{31}
    \langle\rho_k(\mathbf{q})\rangle^t
    =\left(\frac{\beta}{S L}\right)
    \sum\limits_{k',\mathbf{q}'}
    \left[\overline{S}_{k'}(\mathbf{q}')-\overline{\mu}_{k'}(\mathbf{q}';t)\right]
    {\mathfrak M}_{-k',-k}^{\rho\rho}(-\mathbf{q}',-\mathbf{q}),
 \end{equation}
where $\overline{S}_{k}(\mathbf{q})=N_\ion
{S}_{k}(\mathbf{q})w_{k}(\mathbf{q})$.

Denoting $[{\mathfrak
M}^{\rho\rho}]_{k_1,k_2}^{-1}(\mathbf{q}_1,\mathbf{q}_2)$ as the
inverse function of
 ${\mathfrak M}^{\rho\rho}_{k_1,k_2}(\mathbf{q}_1,\mathbf{q}_2)$:

 \begin{equation}\label{31a}
  \sum\limits_{k'',\mathbf{q}''}
  [{\mathfrak M}^{\rho\rho}]_{k,k''}^{-1}(\mathbf{q},\mathbf{q}'')
  {\mathfrak M}^{\rho\rho}_{k'',k'}(\mathbf{q}'',\mathbf{q}')=
  \delta_{k,k'}\delta_{\mathbf{q},\mathbf{q}'}
 \end{equation}
from the Fourier-component of the electrons one can find
the electrochemical potential:
 \begin{equation}\label{32}
    \overline{\mu}_{k}(\mathbf{q};t)=\overline{S}_{k}(\mathbf{q})-
    \left(\frac{\beta}{S L}\right)^{-1}
    \sum\limits_{k',\mathbf{q}'}
    \left\langle\rho_{k'}(\mathbf{q}')\right\rangle^t
    [{\mathfrak M}^{\rho\rho}]_{k',k}^{-1}(-\mathbf{q}',-\mathbf{q}).
 \end{equation}
As we can see, the Fourier-component of the electrochemical
potential in the Gaussian approximation is expressed in terms of
the structure factor of the ionic subsystem and the
Fourier-component of the local pseudopotential of interaction
between electrons and ions. Time-dependence is described with the
average equilibrium value of electrons density, renormalized in
terms of the structure factor of the ionic subsystem, the
pseudopotential $w_k(\mathbf{q})$ and the inverse function
$[{\mathfrak M}^{\rho\rho}]_{k',k}^{-1}(-\mathbf{q}',-\mathbf{q})$
of the pair irreducible cumulant average of electrons density
fluctuations. Similarly we exclude the parameters
$\mathbf{A}_k(\mathbf{q};t)$ from (\ref{30}) using the
thermodynamical relation in the Gaussian approximation:
 \[
  \frac{\delta \Phi^{\mathrm{(G)}}(t)}{\delta \frac{\beta}{SL}\mathbf{A}_k(\mathbf{q};t)}=
  \langle\mathbf{p}_k(\mathbf{q})\rangle^t,
 \]
hence, one can find
 \begin{equation}\label{33}
    \langle\mathbf{p}_k(\mathbf{q})\rangle^t=
    -\left(\frac{\beta}{S L}\right)
    \sum\limits_{k',\mathbf{q}'}
    \mathbf{A}_{k'}(\mathbf{q}')\cdot
    \bm{\mathfrak M}^{\mathbf{p}\mathbf{p}}_{k',k}(-\mathbf{q}',-\mathbf{q}).
 \end{equation}

Defining $[\bm{\mathfrak
M}^{\mathbf{p}\mathbf{p}}]_{-k',-k''}^{-1}(-\mathbf{q}',-\mathbf{q}'')$
as the inverse function of $\bm{\mathfrak
M}^{\mathbf{p}\mathbf{p}}_{-k',-k''}(-\mathbf{q}',-\mathbf{q}'')$:
 \[
  \sum\limits_{k'',\mathbf{q}''}
  [\bm{\mathfrak M}^{\mathbf{p}\mathbf{p}}]_{k,k''}^{-1}(\mathbf{q},\mathbf{q}'')
  \bm{\mathfrak M}^{\mathbf{p}\mathbf{p}}_{k'',k'}(\mathbf{q}'',\mathbf{q}')=
  \delta_{k,k'}\delta_{\mathbf{q},\mathbf{q}'}
 \]
for the Fourier-component of
$\mathbf{A}_k(\mathbf{q};t)$ one can obtain from (\ref{33}):
 \begin{equation}\label{34}
    \mathbf{A}_k(\mathbf{q};t)=-
    \left(\frac{\beta}{S L}\right)^{-1}
    \sum\limits_{k',\mathbf{q}'}
    \left\langle\mathbf{p}_{k'}(\mathbf{q}')\right\rangle^t\cdot
    [\bm{\mathfrak M}^{\mathbf{p}\mathbf{p}}]_{-k',-k}^{-1}(-\mathbf{q}',-\mathbf{q}),
 \end{equation}
the time-dependence of which is described with the average value of a
density momentum operator, renormalized via function $[\bm{\mathfrak
M}^{\mathbf{p}\mathbf{p}}]_{-k',-k}^{-1}(-\mathbf{q}',-\mathbf{q}),$
which is the inverse function of the pair irreducible cumulant
average value of the momentum density fluctuation of electrons. In
consideration of (\ref{32}) and (\ref{34}), the entropy operator
could be written down as follows:
 \begin{eqnarray}\label{35}
      \widehat{S}^{\rm G}(t) & =&\beta\frac{N}{2S}{\sum\limits_{\mathbf{q}}}'\nu(\mathbf{q}|0)+\ln Z_\jell 
        -\sum\limits_{k_1,k_2}
        \sum\limits_{\mathbf{q}_1,\mathbf{q}_2}
        \Big[
         \left\langle\rho_{k_1}(\mathbf{q}_1)\right\rangle^t
                 [{\mathfrak M}^{\rho\rho}]_{-k_1,-k_2}^{-1}(-\mathbf{q}_1,-\mathbf{q}_2)
         \left\langle\rho_{k_2}(\mathbf{q}_2)\right\rangle^t
     \nonumber\\
        &+&\left\langle\mathbf{p}_{k_1}(\mathbf{q}_1)\right\rangle^t\cdot[\bm{\mathfrak M}^{\mathbf{p}\mathbf{p}}]_{-k_1,-k_2}^{-1}(-\mathbf{q}_1,-\mathbf{q}_2)
        \cdot\left\langle\mathbf{p}_{k_2}(\mathbf{q}_2)\right\rangle^t\Big]\nonumber
     \\
   &+&\beta \bigg\{H-\frac1{S L}\sum\limits_{k,\mathbf{q}}
         \bigg[ \overline{S}_{k}(\mathbf{q})-
    \left(\frac{\beta}{S L}\right)^{-1}
    \sum\limits_{k',\mathbf{q}'}
    \left\langle\rho_{k'}(\mathbf{q}')\right\rangle^t
    [{\mathfrak M}^{\rho\rho}]_{-k',-k}^{-1}(-\mathbf{q}',-\mathbf{q})\bigg]\rho_k(\mathbf{q})
      \nonumber \\
      &+&\frac{1}{S L}\left(\frac{\beta}{S L}\right)^{-1}
      \sum\limits_{k,k'}
      \sum\limits_{\mathbf{q},\mathbf{q}'}
      \left\langle\mathbf{p}_{k'}(\mathbf{q}')\right\rangle^t\cdot[\bm{\mathfrak M}^{\mathbf{p}\mathbf{p}}]_{-k',-k}^{-1}(-\mathbf{q}',-\mathbf{q})
        \cdot\mathbf{p}_{k}(\mathbf{q}) \bigg\}   .
 \end{eqnarray}
Then, in order to calculate the nonequilibrium statistical operator
(\ref{5}) in the Gaussian approximation (\ref{29}), (\ref{35}) one
should reveal the structure of the Kawasaki-Gunton projection
operator (\ref{6}), and its effect and the effect of the Liouville
operator on $\rho^{\mathrm{(G)}}_{\mathrm{q}}$ as it is shown in appendix~A. Then,
considering (\ref{38}), (\ref{39}) for the
nonequilibrium statistical operator one can get:
 \begin{eqnarray}\label{43}
    \rho(t) & =& \rho^{\mathrm{(G)}}_{\mathrm{q}}(t)-\sum\limits_{k,\mathbf{q}}
    \int\limits_{-\infty}^t\!\!
    \ee^{\varepsilon(t'-t)}T_{\mathrm{q}}^{\mathrm G}(t,t') \nonumber\\
      &\times&
       \left\{
        \int\limits_0^1\dd\tau
   \left[\rho^{\mathrm{(G)}}_{\mathrm{q}}(t')\right]^\tau
   I_{\rho}^{\mathrm{(G)}}(k,\mathbf{q};t')
   \left[\rho^{\mathrm{(G)}}_{\mathrm{q}}(t')\right]^{1-\tau}
   W^{\mathrm{(G)}}_{\rho\rho}(k,\mathbf{q};t')
   \right.
   \nonumber
       \\
      & + &     \left.
        \int\limits_0^1\dd\tau
   \left[\rho^{\mathrm{(G)}}_{\mathrm{q}}(t')\right]^\tau
   \mathbf{I}_{\mathbf{p}}^{\mathrm{(G)}}(k,\mathbf{q};t')
   \left[\rho^{\mathrm{(G)}}_{\mathrm{q}}(t')\right]^{1-\tau}\cdot
   \mathbf{W}^{\mathrm{(G)}}_{\mathbf{p}\mathbf{p}}(k,\mathbf{q};t')
   \right\}\dd t',
 \end{eqnarray}
where
 \begin{equation}\label{44}
      I_{\rho}^{\mathrm{(G)}}(k,\mathbf{q};t')  = \left[1-{\cal P}^{\mathrm{(G)}}(t')\right]\ii L \rho_k(\mathbf{q}),\qquad
      \mathbf{I}_{\mathbf{p}}^{\mathrm{(G)}}(k,\mathbf{q};t')   = \left[1-{\cal P}^{\mathrm{(G)}}(t')\right]\ii L \mathbf{p}_k(\mathbf{q})
 \end{equation}
are the generalized fluxes, ${\cal P}^{\mathrm{(G)}}(t)$ is a Mori-like
projection operator, which effects as follows:
 \begin{eqnarray}\label{45}
      {\cal P}^{\mathrm{(G)}}(t)\hat{A} & =&
      \sum\limits_{k,\mathbf{q}}
      \sum\limits_{k',\mathbf{q}'}\bigg[
      \delta\rho_{k'}(\mathbf{q}';t)
    [{\mathfrak M}^{\rho\rho}]_{-k',-k''}^{-1}(-\mathbf{q}',-\mathbf{q}'')
        \left\langle\rho_{k''}(\mathbf{q}'')\hat{A}\right\rangle^t_{\mathrm{G}}
\nonumber\\
        &+& \delta\mathbf{p}_{k'}(\mathbf{q}';t)\cdot
    [\bm{\mathfrak
    M}^{\mathbf{p}\mathbf{p}}]_{-k',-k''}^{-1}(-\mathbf{q}',-\mathbf{q}'')\cdot
    \left\langle\mathbf{p}_{k''}(\mathbf{q}'')\hat{A}\right\rangle^t_{\mathrm{G}}\bigg],
 \end{eqnarray}
where
 \begin{eqnarray*}
  \langle\ldots\rangle^t_{\mathrm{G}}&=&\Sp\left[\ldots \rho^{\rm G}_{\mathrm{q}}(t)\right],
\\%
  \delta\rho_{k'}(\mathbf{q}';t)=
  \rho_{k'}(\mathbf{q}')-
          \left\langle\rho_{k'}(\mathbf{q}')\right\rangle^t, &&
  \delta\mathbf{p}_{k'}(\mathbf{q}';t)=
  \mathbf{p}_{k'}(\mathbf{q}')-
          \left\langle\mathbf{p}_{k'}(\mathbf{q}')\right\rangle^t.
 \end{eqnarray*}
Using the effect of operator ${\cal P}^{\mathrm{(G)}}(t')$ on $\ii L
\rho_k(\mathbf{q})$, one can show that
$I_{\rho}^{\mathrm{(G)}}(k,\mathbf{q};t')=0$. According to~(\ref{43}) the
nonequilibrium statistical operator in the Gaussian approximation
is the functional of the observable values
$\langle\rho_k(\mathbf{q})\rangle^t$,
$\langle\mathbf{p}_k(\mathbf{q})\rangle^t$ and the generalized
fluxes of momentum density
$\mathbf{I}_{\mathbf{p}}^{\mathrm{(G)}}(k,\mathbf{q};t')$. One can obtain the
relevant transport equations for
$\langle\rho_k(\mathbf{q})\rangle^t$ and
$\langle\mathbf{p}_k(\mathbf{q})\rangle^t$ with the nonequilibrium
statistical operator as follows:
  \begin{eqnarray}\label{52}
  \frac{\partial}{\partial t}\langle\rho_k(\mathbf{q})\rangle^t  & + &\ii k\mathbf{q}\frac1m\cdot\langle\mathbf{p}_k(\mathbf{q})\rangle^t=0,
\\[4mm]
\frac{\partial}{\partial t}\langle\mathbf{p}_k(\mathbf{q})\rangle^t & + &
    \sum\limits_{k',\mathbf{q}'}
   \big\langle\mathbf{p}_k(\mathbf{q})\dot{\rho}_{k'}(\mathbf{q}')\big\rangle^t_{\mathrm{G}}
    W_{\rho\rho}^{\mathrm{(G)}}(k',\mathbf{q}';t)\nonumber\\
    \label{53}
    & - &
    \sum\limits_{k,\mathbf{q}}
    \int\limits_{-\infty}^t\!\!\ee^{\varepsilon(t'-t)}
    \varphi_{\mathbf{p}\mathbf{p}}^{\mathrm{(G)}}(k,\mathbf{q};k',\mathbf{q}';t,t')
    \mathbf{W}_{\mathbf{p}\mathbf{p}}^{\mathrm{(G)}}(k',\mathbf{q}';t')\dd t'  =0 ,
  \end{eqnarray}
where
  \begin{equation}\label{49}
    \varphi_{\mathbf{p}\mathbf{p}}^{\mathrm{(G)}}(k,\mathbf{q};k',\mathbf{q}';t,t')
    =
    \Sp\!
    \left\{
       \mathbf{I}_{\mathbf{p}}^{\mathrm{(G)}}(k,\mathbf{q};t) \,
       T_{\mathrm{q}}^{\rm G}(t,t')\!\!
        \int\limits_0^1\!\!\dd\tau\!
   \left[\rho^{\mathrm{(G)}}_{\mathrm{q}}(t')\right]^\tau \!\!
   \mathbf{I}_{\mathbf{p}}^{\mathrm{(G)}}(k',\mathbf{q}';t')\!
   \left[\rho^{\mathrm{(G)}}_{\mathrm{q}}(t')\right]^{1-\tau}\!
    \right\}
  \end{equation}
is the generalized memory function that describes dissipative
processes and takes into account
  \begin{equation}\label{51}
   \big\langle\dot{\mathbf{p}}_k(\mathbf{q})\big\rangle^t_{\mathrm{G}}=
    -\sum\limits_{k',\mathbf{q}'}
   \big\langle\mathbf{p}_k(\mathbf{q})\dot{\rho}_{k'}(\mathbf{q}')\big\rangle^t_{\mathrm{G}}
     W_{\rho\rho}^{\mathrm{(G)}}(k',\mathbf{q}';t)    .
  \end{equation}
It is the time correlation function of the generalized fluxes of
momentum density, averaged with the quasi-equilibrium statistical
operator $\rho_{\mathrm{q}}^{\mathrm{(G)}}(t)$ in the Gaussian approximation.
$\dot{\mathbf{p}}_k(\mathbf{q})=\ii L_N
{\mathbf{p}}_k(\mathbf{q})=\ii k
\mathbf{q}:\mathop{\mathbf{T}}\limits^{\leftrightarrow}{\!\!}_k(\mathbf{q})$
denotes the tensor operator of a viscous stress of the electron
subsystem. This means that the generalized memory function
(\ref{49}) defines the generalized viscosity coefficient of the
electron subsystem:
  \begin{equation}\label{50}
    \varphi_{\mathbf{p}\mathbf{p}}^{\mathrm{(G)}}(k,\mathbf{q};k',\mathbf{q}';t,t')
    =-k\mathbf{q}:\eta^{\mathrm{(G)}}(k,\mathbf{q};k',\mathbf{q}';t,t''): k'\mathbf{q}'.
  \end{equation}

It is notable that the system of equations (\ref{52}), (\ref{53})
has the same structure as in the case of weakly nonequilibrium
processes with the only difference in the relevant averages:
  $\rho_{\mathrm{q}}^{\mathrm{(G)}}(t)\to\rho_{\mathrm{q}}^{0}(t),$ where
   \begin{eqnarray*}
    \rho_{\mathrm{q}}^{0}(t) & = & \ \rho_0
    \left\{
     1
     +
     \sum\limits_{k,\mathbf{q}}\sum\limits_{k',\mathbf{q}'}
     \left[
      \delta\rho_{k'}(\mathbf{q}',t)
      [\widetilde{{\mathfrak M}}^{\rho\rho}]^{-1}_{-k',-k}(-\mathbf{q},-\mathbf{q}')
      \int\limits_0^1\!\!\dd\tau\rho_0^\tau\rho_k(\mathbf{q})\rho_0^{-\tau}\right.\right. \\
      & + & \left.\left.
      \delta\mathbf{p}_{k'}(\mathbf{q}',t)\cdot
      [\widetilde{\bm{\mathfrak M}}^{\mathbf{p}\mathbf{p}}]^{-1}_{-k',-k}(-\mathbf{q},-\mathbf{q}')
      \int\limits_0^1\!\!\dd\tau\rho_0^\tau\cdot\mathbf{p}_k(\mathbf{q})\rho_0^{-\tau}
     \right]
    \right\},
\end{eqnarray*}
where $\rho_0$ is the equilibrium statistical operator of the
system.

In the next section we work with the following higher
approximation for the quasi-equilibrium partition function, when
the static correlations between the operators of electron density
and electron momentum density occur with the third-order cumulant
averages.

\section{The quasi-equilibrium partition function in the
approximation of the third-order cumulant averages}

According to (\ref{16}), (\ref{21}) the relevant statistical
operator $\rho_{\mathrm{q}}(t)$ in this approximation is
  \begin{equation}\label{54}
    \rho_{\mathrm{q}}^{\rm G+1}(t)=\exp
    \left[-\hat{S}^{(\rm G+1)}(t)\right],
  \end{equation}
  \begin{eqnarray}\label{55a}
\hat{S}^{(\rm G+1)}(t)
& = & \beta\frac{N}{2S}{\sum_{\mathbf{q}}}'\nu(\mathbf{q}|0)+\ln Z_\jell \nonumber\\
& + & \beta\left\{
     H-\frac1{S L}\sum\limits_{k,\mathbf{q}}
           \Big[
             \overline{\mu}_k(\mathbf{q};t)\rho_k(\mathbf{q})+
             \mathbf{A}_k(\mathbf{q};t)\cdot\mathbf{p}_k(\mathbf{q})
           \Big]
            \right\}
      -\Phi^{(\rm G+1)}(t),
  \end{eqnarray}
  \begin{eqnarray}\label{55b}
      \Phi^{(\rm G+1)}(t) & = &\frac12\left(\frac{\beta}{S L}\right)^{2}\sum_{k_1,k_2}\sum_{\mathbf{q}_1,\mathbf{q}_2}
       \left[
        B_{k_1}(\mathbf{q}_1;t){\mathfrak M}^{\rho\rho}_{k_1,k_2}(\mathbf{q}_1,\mathbf{q}_2)B_{k_2}(\mathbf{q}_2;t)\right. \nonumber\\
        & + & \left.
        \mathbf{A}_{k_1}(\mathbf{q}_1;t)\cdot\bm{\mathfrak M}^{\mathbf{p}\mathbf{p}}_{k_1,k_2}(\mathbf{q}_1,\mathbf{q}_2)\cdot\mathbf{A}_{k_2}(\mathbf{q}_2;t)
       \right] \nonumber \\
        & + & \frac{\ii}{3!}\left(\frac{\beta}{S L}\right)^{3}\sum_{k_1,k_2,k_3}\sum_{\mathbf{q}_1,\mathbf{q}_2,\mathbf{q}_3}
       \left[
        B_{k_1}(\mathbf{q}_1;t)
        B_{k_2}(\mathbf{q}_2;t)
        B_{k_3}(\mathbf{q}_3;t)
        {\mathfrak M}^{\rho\rho\rho}_{k_1,k_2,k_3}(\mathbf{q}_1,\mathbf{q}_2,\mathbf{q}_3)
        \right. \nonumber \\
        & + & \left.
        3\mathbf{A}_{k_1}(\mathbf{q}_1;t)\cdot\bm{\mathfrak M}^{\mathbf{p}\mathbf{p}\rho}_{k_1,k_2,k_3}(\mathbf{q}_1,\mathbf{q}_2,\mathbf{q}_3)\cdot\mathbf{A}_{k_2}(\mathbf{q}_2;t)
        B_{k_3}(\mathbf{q}_3;t)
       \right].
  \end{eqnarray}
In order to obtain an explicit form of the nonequilibrium
statistical operator and the transport equations within the
approximation (\ref{54}) one should exclude parameters
$\overline{\mu}_k(\mathbf{q};t)$ and $\mathbf{A}_k(\mathbf{q};t)$
from (\ref{55a}). In the same way as in the Gaussian approximation
case we use the thermodynamical relations:
\begin{equation}\label{56}
\frac{\delta\Phi^{(\mathrm G+1)}(t)}{\delta\frac{\beta}{SL}\overline{\mu}_k(\mathbf{q};t)}=
\langle\rho_k(\mathbf{q})\rangle^t,
\end{equation}
\begin{equation}\label{57}
\frac{\delta\Phi^{(\mathrm G+1)}(t)}{\delta\frac{\beta}{SL}
\mathbf{A}_k(\mathbf{q};t)}=\langle
\mathbf{p}_k(\mathbf{q})\rangle^t.
 \end{equation}
Considering the structure of $\Phi^{(\mathrm G+1)}(t)$ in
$\hat{S}^{(\mathrm G+1)}(t)$ from (\ref{56}) one can get equations to
define the electrochemical potential of the electron
subsystem of a semi-bounded metal within the generalized
``jellium'' model:
\begin{eqnarray}\label{58}
      \left\langle\rho_{k}(\mathbf{q})\right\rangle^t & = & \frac{\beta}{S L}
      \sum_{k_1,\mathbf{q}_1}
        B_{k_1}(\mathbf{q}_1;t){\mathfrak M}^{\rho\rho}_{k_1,k}(\mathbf{q}_1,\mathbf{q}) \nonumber \\
        & + & \frac{\ii}{2!}\left(\frac{\beta}{S L}\right)^{2}
           \sum_{k_1,k_2}\sum_{\mathbf{q}_1,\mathbf{q}_2}
           \left[
              B_{k_1}(\mathbf{q}_1;t)
              {\mathfrak M}^{\rho\rho\rho}_{k_1,k_2,k}(\mathbf{q}_1,\mathbf{q}_2,\mathbf{q})
              B_{k_2}(\mathbf{q}_2;t)
           \right. \nonumber \\
        & - &  \left. 3\mathbf{A}_{k_1}(\mathbf{q}_1;t)\cdot
              \bm{\mathfrak M}^{\mathbf{p}\mathbf{p}\rho}_{k_1,k_2,k}(\mathbf{q}_1,\mathbf{q}_2,\mathbf{q})
              \cdot\mathbf{A}_{k_2}(\mathbf{q}_2;t)
           \right],
\end{eqnarray}
and to define $\mathbf{A}_{k}(\mathbf{q};t)$ from (\ref{57})
one can find:
\begin{eqnarray}\label{59}
      \left\langle\mathbf{p}_{k}(\mathbf{q})\right\rangle^t & = & -\frac{\beta}{S L}
      \sum_{k_1,\mathbf{q}_1}
        \mathbf{A}_{k_1}(\mathbf{q}_1;t)\cdot\bm{\mathfrak M}^{\mathbf{p}\mathbf{p}}_{k_1,k}(\mathbf{q}_1,\mathbf{q})\nonumber \\
        & - & \ii\left(\frac{\beta}{S L}\right)^{2}
           \sum_{k_1,k_2}\sum_{\mathbf{q}_1,\mathbf{q}_2}
              \mathbf{A}_{k_1}(\mathbf{q}_1;t)\cdot
              \bm{\mathfrak M}^{\mathbf{p}\mathbf{p}\rho}_{k_1,k_2,k}(\mathbf{q}_1,\mathbf{q}_2,\mathbf{q})
              B_{k_2}(\mathbf{q}_2;t).
\end{eqnarray}
The detailed calculations are presented in the appendix~B. Considering~(\ref{71}) the nonequilibrium statistical operator of the electron
subsystem of a semi-bounded metal within the generalized ``jellium''
model, according to (\ref{5}) we present as follows:
 \begin{eqnarray}\label{75}
    \rho(t) & = & \rho_{\mathrm{q}}^{(\mathrm G+1)}(t) -\sum_{k,\mathbf{q}}\int\limits_{-\infty}^t\!\!\!\!\ee^{\varepsilon(t'-t)}
    T_{\mathrm{q}}^{(\mathrm G+1)}(t,t')\left[1-\mathcal{P}_{\mathrm{q}}^{(\mathrm G+1)}(t')\right]\nonumber \\
      & \times & \Bigg\{ \int\limits_0^1\dd\tau
   \left[\rho^{(\mathrm G+1)}_{\mathrm{q}}(t')\right]^\tau
   \dot{\rho}_{k}(\mathbf{q})
   \left[\rho^{(\mathrm G+1)}_{\mathrm{q}}(t')\right]^{1-\tau} W^{(\mathrm G+1)}_{\rho\rho}(k,\mathbf{q};t')   \nonumber   \\
      & + & \int\limits_0^1\dd\tau
   \left[\rho^{(\mathrm G+1)}_{\mathrm{q}}(t')\right]^\tau
   \dot{\mathbf{p}}_{k}(\mathbf{q})
   \left[\rho^{(\mathrm G+1)}_{\mathrm{q}}(t')\right]^{1-\tau}\cdot \mathbf{W}^{(\mathrm G+1)}_{\mathbf{p}\mathbf{p}}(k,\mathbf{q};t)
   \Bigg\}\dd t',
 \end{eqnarray}
 where
 \begin{equation}\label{42}
    \dot{\rho}_k(\mathbf{q})=\ii L_N \rho_k(\mathbf{q}),\qquad
    \dot{\mathbf{p}}_k(\mathbf{q})=\ii L_N \mathbf{p}_k(\mathbf{q}).
\end{equation}

Then, taking into account the above found we obtain the
transport equations for the reduced description parameters
 $\left\langle\rho_{k}(\mathbf{q})\right\rangle^t$, $\left\langle\mathbf{p}_{k}(\mathbf{q})\right\rangle^t$:
  \begin{eqnarray}\label{76}
    \!\!\frac{\partial}{\partial t}\langle\rho_k(\mathbf{q})\rangle^t\!\! & = & \!\!\langle\dot{\rho}_k(\mathbf{q})\rangle^t_{(\mathrm G+1)} 
    -\sum_{k',\mathbf{q}'}\int\limits_{-\infty}^t\!\!\!\!\ee^{\varepsilon(t'-t)}
    \Phi_{\rho\rho}^{(\mathrm G+1)}(k,\mathbf{q},k',\mathbf{q}';t,t')W^{(\mathrm G+1)}_{\rho\rho}(k,\mathbf{q};t')\dd t' \nonumber \\
    & - & \!\!\sum_{k',\mathbf{q}'}\int\limits_{-\infty}^t\!\!\!\!\ee^{\varepsilon(t'-t)}
    {\Phi}_{\rho\,\mathbf{p}}^{(\mathrm G+1)}(k,\mathbf{q},k',\mathbf{q}';t,t')\cdot \mathbf{W}^{(\mathrm G+1)}_{\mathbf{p}\mathbf{p}}(k,\mathbf{q};t')
    \dd t',
  \end{eqnarray}
  \begin{eqnarray}\label{77}
   \!\! \frac{\partial}{\partial t}\langle\mathbf{p}_k(\mathbf{q})\rangle^t\!\!
    & = & \!\!\langle\dot{\mathbf{p}}_k(\mathbf{q})\rangle^t_{(\mathrm G+1)}
    -\sum_{k',\mathbf{q}'}\int\limits_{-\infty}^t\!\!\!\!\ee^{\varepsilon(t'-t)}
    {\Phi}_{\mathbf{p}\,\rho}^{(\mathrm G+1)}(k,\mathbf{q},k',\mathbf{q}';t,t')W^{(\mathrm G+1)}_{\rho\rho}(k,\mathbf{q};t')\dd t'\nonumber \\
    & - & \!\!\sum_{k',\mathbf{q}'}\int\limits_{-\infty}^t\!\!\!\!\ee^{\varepsilon(t'-t)}
    {\Phi}_{\mathbf{p}\mathbf{p}}^{(\mathrm G+1)}(k,\mathbf{q},k',\mathbf{q}';t,t'): \mathbf{W}^{(\mathrm G+1)}_{\mathbf{p}\mathbf{p}}(k,\mathbf{q};t')
    \dd t',
  \end{eqnarray}
where
\[\langle(\ldots)\rangle_{(\mathrm G+1)}^t=\Sp(\ldots)\rho_{\mathrm{q}}^{(\mathrm G+1)}(t), \]
$\Phi_{\rho\rho}^{(\mathrm G+1)}(k,\mathbf{q},k',\mathbf{q}';t,t')$,
$ {\Phi}_{\rho\,\mathbf{p}}^{(\mathrm G+1)}(k,\mathbf{q},k',\mathbf{q}';t,t')$,
${\Phi}_{\mathbf{p}\,\rho}^{(\mathrm G+1)}(k,\mathbf{q},k',\mathbf{q}';t,t')$,
${\Phi}_{\mathbf{p}\mathbf{p}}^{(\mathrm G+1)}(k,\mathbf{q},k',\mathbf{q}';t,t')$ \linebreak
are the generalized transport kernels that describe diffusive,
visco-diffusive and viscous processes of the electron subsystem of
a semi-bounded metal within the generalized ``jellium'' model.
Particularly,
$\Phi_{\rho\rho}^{(\mathrm G+1)}(k,\mathbf{q},k',\mathbf{q}';t,t')$ and
${\Phi}_{\mathbf{p}\mathbf{p}}^{(\mathrm G+1)}(k,\mathbf{q},k',\mathbf{q}';t,t')$
have the following structure:
  \begin{eqnarray}\label{78}
      \Phi_{\rho\rho}^{(\mathrm G+1)}(k,\mathbf{q},k',\mathbf{q}';t,t')
      & = & \Big\langle
      \dot{\rho}_k(\mathbf{q})
      T_{\mathrm{q}}^{(\mathrm G+1)}(t,t')\left[1-{\cal P}_{\mathrm{q}}^{(\mathrm G+1)}(t')\right] \nonumber \\
      & \times & \int\limits_0^1\dd\tau
   \left[\rho^{(\mathrm G+1)}_{\mathrm{q}}(t')\right]^\tau
   \dot{\rho}_{k'}(\mathbf{q}')
   \left[\rho^{(\mathrm G+1)}_{\mathrm{q}}(t')\right]^{-\tau}\Big\rangle_{(\mathrm G+1)}^{t'} \nonumber \\
   & = &  k\, \mathbf{q}\cdot D^{(\mathrm G+1)}(k,\mathbf{q},k',\mathbf{q}';t,t')\cdot k'\, \mathbf{q}',
  \end{eqnarray}
  \begin{eqnarray}\label{79}
      {\Phi}_{\mathbf{p}\mathbf{p}}^{(\mathrm G+1)}(k,\mathbf{q},k',\mathbf{q}';t,t')
      & = & \Big\langle
      \dot{\mathbf{p}}_k(\mathbf{q})
      T_{\mathrm{q}}^{(\mathrm G+1)}(t,t')\left[1-{\cal P}_{\mathrm{q}}^{(\mathrm G+1)}(t')\right] \nonumber \\
        & \times & \int\limits_0^1\dd\tau
   \left[\rho^{(\mathrm G+1)}_{\mathrm{q}}(t')\right]^\tau
   \dot{\mathbf{p}}_{k'}(\mathbf{q}')
   \left[\rho^{(\mathrm G+1)}_{\mathrm{q}}(t')\right]^{-\tau}\Big\rangle_{(\mathrm G+1)}^{t'} \nonumber \\
   & = &  k\, \mathbf{q}: \eta^{(\mathrm G+1)}(k,\mathbf{q},k',\mathbf{q}';t,t'): k'\, \mathbf{q}',
  \end{eqnarray}
$D^{(\mathrm G+1)}(k,\mathbf{q},k',\mathbf{q}';t,t')$, $\eta^{(\mathrm G+1)}(k,\mathbf{q},k',\mathbf{q}';t,t')$ denote the
coefficients of nonlinear diffusion and nonlinear viscosity of the
electron subsystem of a semi-bounded metal within the ``jellium''
model. The generalized transport equations (\ref{76}), (\ref{77})
and the nonequilibrium statistical operator are strongly nonlinear
ones in comparison with the transport equations (\ref{52}),
(\ref{53}), which correspond to the Gaussian approximation for
$\rho_{\mathrm{q}}^{(\mathrm G)}(t)$.

\section{Conclusion}
Viscoelastic processes in the electron subsystem of a semi-bounded
metal are described on the basis of the generalized ``jellium''
model with the use of the NSO method, where the parameters of the
reduced description are the nonequilibrium average values of
electron density and electron momentum. Applying the functional
integration technique, we have calculated the quasi-equilibrium
partition function for such a system in the case of the model
pseudopotential of electron-ion interaction in a metal in the
Gaussian approximation and in the following higher approximation,
where the static correlations between the operators of electron
density and electron momentum density are taken into account with
the third-order cumulant averages.

We have also obtained the expressions for a nonequilibrium
statistical operator, which enables us to go beyond the linear
approximation with respect to the gradients of electrochemical
potential and average electron density. In the respective
approximations for a nonequilibrium statistical operator we have
derived the generalized transport equations for a nonequilibrium
average values of the electron density operator and the electron
momentum density operator that can be applied to the description
of strongly nonequilibrium processes for an electron subsystem of
a semi-bounded metal. The generalized transport coefficients (that
related, for example, with the generalized viscosity coefficient of
the electron subsystem of a semi-bounded metal) which are
contained in the corresponding transport equations are calculated
using the quasi-equilibrium statistical operator in the respective
approximations: the Gaussian one~(\ref{29}) and the following
higher approximation~(\ref{54}). An important point in such an
approach is that the time correlation functions and the
generalized transport coefficients are calculated with the
quasi-equilibrium statistical operator in the corresponding
approximation and represent the functionals of the observable
quantities $\langle\rho_k(\vec q)\rangle^t$,
$\langle\mathbf{p}_k(\mathbf{q})\rangle^t$ of a certain order. Of
special interest in this approach are the investigations of the
dynamic structure factor for the nonequilibrium electron subsystem
of a semi-bounded metal.

\appendix

\section{Kawasaki-Gunton projection operator
in the Gaussian \\ approximation for $\rho_{\mathrm{q}}^{(\mathrm G)}(t)$}
 \renewcommand{\theequation}{A.\arabic{equation}}
 \setcounter{equation}{0}

Considering (\ref{29}), (\ref{35}) and
\begin{eqnarray}\label{36}
      \frac{\delta\rho^{\mathrm{(G)}}_{\mathrm{q}}(t)}{\delta\left\langle\rho_{k}(\mathbf{q})\right\rangle^t}  & = &
      -
      \sum\limits_{k',\mathbf{q}'}\bigg[
      \rho_{k'}(\mathbf{q}';\tau;t)
 -    \left\langle\rho_{k'}(\mathbf{q}')\right\rangle^t       \bigg]
    [{\mathfrak M}^{\rho\rho}]_{-k',-k}^{-1}(-\mathbf{q}',-\mathbf{q})
    \rho^{\mathrm{(G)}}_{\mathrm{q}}(t),\\
\label{37}
      \frac{\delta\rho^{\mathrm{(G)}}_{\mathrm{q}}(t)}{\delta\left\langle\mathbf{p}_{k}(\mathbf{q})\right\rangle^t} %
     & = &
      -
      \sum\limits_{k',\mathbf{q}'}\bigg[
      \mathbf{p}_{k'}(\mathbf{q}';\tau;t)
        -   \left\langle\mathbf{p}_{k'}(\mathbf{q}')\right\rangle^t
    \bigg]\cdot
    [\bm{\mathfrak M}^{\mathbf{p}\mathbf{p}}]_{-k',-k}^{-1}(-\mathbf{q}',-\mathbf{q})
    \rho^{\mathrm{(G)}}_{\mathrm{q}}(t),
 \end{eqnarray}
where
 \[
   \rho_{k'}(\mathbf{q}';\tau;t)=\int\limits_0^1\dd\tau
   \left[\rho^{\mathrm{(G)}}_{\mathrm{q}}(t)\right]^\tau
   \rho_{k'}(\mathbf{q}')
   \left[\rho^{\mathrm{(G)}}_{\mathrm{q}}(t)\right]^{-\tau},
 \]
 \[
   \mathbf{p}_{k'}(\mathbf{q}';\tau;t)=\int\limits_0^1\dd\tau
   \left[\rho^{\mathrm{(G)}}_{\mathrm{q}}(t)\right]^\tau
   \mathbf{p}_{k'}(\mathbf{q}')
   \left[\rho^{\mathrm{(G)}}_{\mathrm{q}}(t)\right]^{-\tau},
 \]
we find the Kawasaki-Gunton projection operator:
 \begin{eqnarray}\label{38}
\lefteqn{      P^{\mathrm{(G)}}_{\mathrm{q}}(t)\rho'
      = \left\{\rho^{\mathrm{(G)}}_{\mathrm{q}}(t)+
      \sum\limits_{k,\mathbf{q}}
      \sum\limits_{k',\mathbf{q}'}
      \left[
      \rho_{k'}(\mathbf{q}';\tau,t)
  -    \left\langle\rho_{k'}(\mathbf{q}')\right\rangle^t   \right]\right.}
 \nonumber  \\
 &&\mbox{} \times
    [{\mathfrak M}^{\rho\rho}]_{-k',-k}^{-1}(-\mathbf{q}',-\mathbf{q})
    \left\langle\rho_{k}(\mathbf{q})\right\rangle^t
    \rho^{\mathrm{(G)}}_{\mathrm{q}}(t)
    \Sp(\rho') \nonumber
 \\
 &&\mbox{} +\!
      \sum\limits_{k,\mathbf{q}}
      \sum\limits_{k',\mathbf{q}'}
     \left(
      \mathbf{p}_{k'}(\mathbf{q}';\tau,t)
      -
      \left\langle\mathbf{p}_{k'}(\mathbf{q}')\right\rangle^t
    \right)\cdot
    [\bm{\mathfrak M}^{\mathbf{p}\mathbf{p}}]_{-k',-k}^{-1}(-\mathbf{q}',-\mathbf{q})
    \cdot\left\langle\mathbf{p}_{k}(\mathbf{q})\right\rangle^t
    \rho^{\mathrm{(G)}}_{\mathrm{q}}(t)
    \Sp(\rho')
\nonumber  \\
 &&\mbox{} -\!
      \sum\limits_{k,\mathbf{q}}
      \sum\limits_{k',\mathbf{q}'}
    \left(
      \rho_{k'}(\mathbf{q}';\tau,t)
-    \left\langle\rho_{k'}(\mathbf{q}')\right\rangle^t
    \right)
    [{\mathfrak M}^{\rho\rho}]_{-k',-k}^{-1}(-\mathbf{q}',-\mathbf{q})
    \Sp\left(\rho_{k}(\mathbf{q})\rho'\right)    \rho^{\mathrm{(G)}}_{\mathrm{q}}(t)
\nonumber \\
 &&\mbox{} -\!
      \left.\sum\limits_{k,\mathbf{q}}
      \sum\limits_{k',\mathbf{q}'}
    \left(
      \mathbf{p}_{k'}(\mathbf{q}';\tau,t)
      -
      \left\langle\mathbf{p}_{k'}(\mathbf{q}')\right\rangle^t
    \right)\cdot
    [\bm{\mathfrak M}^{\mathbf{p}\mathbf{p}}]_{-k',-k}^{-1}(-\mathbf{q}',-\mathbf{q})
    \cdot\Sp(\mathbf{p}_{k}(\mathbf{q})\rho')\right\}
    \rho^{\mathrm{(G)}}_{\mathrm{q}}(t).
 \end{eqnarray}

Taking into account:
\begin{eqnarray}\label{39}
   \ii L_N \rho^{\mathrm{(G)}}_{\mathrm{q}}(t)
   & = & \sum\limits_{k,\mathbf{q}}W^{\mathrm{(G)}}_{\rho\rho}(k,\mathbf{q};t)
   \int\limits_0^1\dd\tau
   \left[\rho^{\mathrm{(G)}}_{\mathrm{q}}(t)\right]^\tau
   \dot{\rho}_{k}(\mathbf{q})
   \left[\rho^{\mathrm{(G)}}_{\mathrm{q}}(t)\right]^{-\tau} \nonumber \\
   & - & \sum\limits_{k,\mathbf{q}}\mathbf{W}^{\mathrm{(G)}}_{\mathbf{p}\mathbf{p}}(k,\mathbf{q};t)
   \int\limits_0^1\dd\tau
   \left[\rho^{\mathrm{(G)}}_{\mathrm{q}}(t)\right]^\tau
   \cdot\dot{\mathbf{p}}_{k}(\mathbf{q})
   \left[\rho^{\mathrm{(G)}}_{\mathrm{q}}(t)\right]^{-\tau},
\end{eqnarray}
where
 \begin{eqnarray}\label{40}
      W^{\mathrm{(G)}}_{\rho\rho}(k,\mathbf{q};t) & = &\frac{\beta}{S L}\overline{\mu}_k(\mathbf{q};t) \nonumber \\
        & = &\frac{\beta}{S L}
        \left[
         \overline{S}_k(\mathbf{q})-
         \left(\frac{\beta}{S L}\right)^{-1}
          \sum\limits_{k',\mathbf{q}'}
          \left\langle\rho_{k'}(\mathbf{q}')\right\rangle^t
             \left[{\mathfrak M}^{\rho\rho}\right]_{-k',-k}^{-1}(-\mathbf{q}',-\mathbf{q})
        \right],\\
 \label{41}
      \mathbf{W}^{\mathrm{(G)}}_{\mathbf{p}\mathbf{p}}(k,\mathbf{q};t) & = &
                   \sum\limits_{k',\mathbf{q}'}
          \left\langle\mathbf{p}_{k'}(\mathbf{q}')\right\rangle^t\cdot
             \left[\bm{\mathfrak
             M}^{\mathbf{p}\mathbf{p}}\right]_{-k',-k}^{-1}(-\mathbf{q}',-\mathbf{q}).
 \end{eqnarray}

\section{Approximate determination of $\bar{\mu}_{k}(\mathbf{q};t)$ in $\rho_{\mathrm{q}}^{(\mathrm G)}(t)$}
 \renewcommand{\theequation}{B.\arabic{equation}}
 \setcounter{equation}{0}

Then, we apply the approximation~\cite{KMVT2011} that linearize the
equations~(\ref{58}), (\ref{59}) with values
$\overline{\mu}_k(\mathbf{q};t)$ (\ref{33}) and
$\mathbf{A}_k(\mathbf{q};t)$ (\ref{35}) in the Gaussian
approximation. So in (\ref{59}) one can obtain:
\begin{eqnarray}\label{60}
      \left\langle\mathbf{p}_{k}(\mathbf{q})\right\rangle^t & = & -\frac{\beta}{S L}
      \sum_{k_1,\mathbf{q}_1}
        \mathbf{A}_{k_1}(\mathbf{q}_1;t)\cdot
        G^{(3)}_{k_1,k}(\mathbf{q}_1,\mathbf{q};t),\\
\label{61}
    \mathbf{G}^{(3)}_{k_1,k}(\mathbf{q}_1,\mathbf{q};t) & = &
    \bm{\mathfrak M}^{\mathbf{p}\mathbf{p}}_{k_1,k}(\mathbf{q}_1,\mathbf{q}) \nonumber \\[2mm]
      & + & \ii
           \sum_{k',k_2}\sum_{\mathbf{q}',\mathbf{q}_2}
           \left\langle\rho_{k'}(\mathbf{q}';t)\right\rangle^t
              \left[{\mathfrak M}^{\rho\rho}\right]^{-1}_{k',k_2}(\mathbf{q}',\mathbf{q}_2)
              \bm{\mathfrak M}^{\mathbf{p}\mathbf{p}\rho}_{k_1,k_2,k}(\mathbf{q}_1,\mathbf{q}_2,\mathbf{q}).
\end{eqnarray}
Defining the inverse function
$\left[\mathbf{G}^{(3)}\right]^{-1}_{k_1,k}(\mathbf{q}_1,\mathbf{q};t)$ of
 $\mathbf{G}^{(3)}_{k_1,k}(\mathbf{q}_1,\mathbf{q};t)$:
 \[
  \sum_{k'',\mathbf{q}''}
  \left[\mathbf{G}^{(3)}\right]^{-1}_{k',k''}(\mathbf{q}',\mathbf{q}'';t)
  \mathbf{G}^{(3)}_{k'',k_1}(\mathbf{q}'',\mathbf{q}_1;t)=
  \delta_{k',k_1}\delta_{\mathbf{q}',\mathbf{q}_1}\,,
 \]
we find from (\ref{60}) the following:
 \begin{equation}\label{62}
    \mathbf{A}_{k_1}(\mathbf{q}_1;t)=
    -\left(\frac{\beta}{S L}\right)^{-1}
    \sum_{k',\mathbf{q}'}
    \left\langle\mathbf{p}_{k'}(\mathbf{q}')\right\rangle^t\cdot
    \left[\mathbf{G}^{(3)}\right]^{-1}_{k',k_1}(\mathbf{q}',\mathbf{q}_1;t).
 \end{equation}
Considering (\ref{62}) with the linearization (\ref{58}), we
obtain the equation for defining $\overline{\mu}_k(\mathbf{q};t)$:
\begin{equation}\label{63}
    \begin{split}
      \left\langle\rho_{k}(\mathbf{q})\right\rangle^t & =\frac{\beta}{S L}
      \sum_{k',\mathbf{q}'}
        \left(
          \overline{S}_{k'}(\mathbf{q}')-\overline{\mu}_{k'}(\mathbf{q}';t)
        \right)
          {\mathfrak M}^{\rho\rho}_{k',k}(\mathbf{q}',\mathbf{q}) \\
        & +{\ii}\frac{\beta}{S L}
           \sum_{k_1,k_2,k'}\sum_{\mathbf{q}_1,\mathbf{q}_2,\mathbf{q}'}
           \!\!\left\langle\rho_{k'}(\mathbf{q}')\right\rangle^t
           \left[{\mathfrak M}^{\rho\rho}\right]^{-1}_{k',k_1}(\mathbf{q}',\mathbf{q}_1)
        \left(
          \overline{S}_{k_2}(\mathbf{q}_2)-\overline{\mu}_{k_2}(\mathbf{q}_2;t)
        \right)
              {\mathfrak M}^{\rho\rho\rho}_{k_1,k_2,k}(\mathbf{q}_1,\mathbf{q}_2,\mathbf{q}) \\
        &   -\frac{3\ii}{2!}
            \sum_{k',k''}\sum_{\mathbf{q}',\mathbf{q}''}
             \left\langle\mathbf{p}_{k'}(\mathbf{q'})\right\rangle^t\cdot
              \mathbf{K}^{\mathbf{p}\mathbf{p}\rho}_{k',k'',k}(\mathbf{q}',\mathbf{q}'',\mathbf{q})
              \cdot\left\langle\mathbf{p}_{k''}(\mathbf{q''})\right\rangle^t,
      \end{split}
\end{equation}
where
 \begin{equation}\label{64}
    \mathbf{K}^{\mathbf{p}\mathbf{p}\rho}_{k',k'',k}(\mathbf{q}',\mathbf{q}'',\mathbf{q})
    =\sum_{k_1,\mathbf{q}_1}
     \sum_{k_2,\mathbf{q}_2}
     \left[\mathbf{G}^{(3)}\right]^{-1}_{k',k_1}(\mathbf{q}',\mathbf{q}_1;t)
     \bm{\mathfrak M}^{\mathbf{p}\mathbf{p}\rho}_{k_1,k_2,k}(\mathbf{q}_1,\mathbf{q}_2,\mathbf{q})
     \left[\mathbf{G}^{(3)}\right]^{-1}_{k_2,k''}(\mathbf{q}_2,\mathbf{q}'';t)
 \end{equation}
is the function that takes into account the complicated
renormalization of the cumulant average
``momentum--momentum--density''. Then, selecting terms from
$\overline{\mu}_k(\mathbf{q};t)$, we rewrite (\ref{63}):
\begin{eqnarray}\label{65}
      \left\langle\rho_{k}(\mathbf{q})\right\rangle^t & = & \frac{\beta}{S L}G_k^{(1)}(\mathbf{q})-
       \frac{\beta}{S L}
       \sum_{k',\mathbf{q}'}\overline{\mu}_{k'}(\mathbf{q}';t)G_{k',k}^{(2)}(\mathbf{q}',\mathbf{q};t) \nonumber       \\
        & - & \frac{3\ii}{2!}
            \sum_{k',k''}\sum_{\mathbf{q}',\mathbf{q}''}
             \left\langle\mathbf{p}_{k'}(\mathbf{q'})\right\rangle^t\cdot
              \mathbf{K}^{\mathbf{p}\mathbf{p}\rho}_{k',k'',k}(\mathbf{q}',\mathbf{q}'',\mathbf{q})
              \cdot\left\langle\mathbf{p}_{k''}(\mathbf{q''})\right\rangle^t,
\end{eqnarray}
where
 \begin{eqnarray}\label{66}
    G_k^{(1)}(\mathbf{q}) & = & \sum_{k',\mathbf{q}'}
    \Big[
     \overline{S}_{k'}(\mathbf{q}')
     {\mathfrak M}^{\rho\rho}_{k',k}(\mathbf{q}',\mathbf{q})   \nonumber
     \\
      & + & \ii
      \sum_{k_1,\mathbf{q}_1}\sum_{k_2,\mathbf{q}_2}
      \left\langle\rho_{k'}(\mathbf{q}')\right\rangle^t
      \left[{\mathfrak M}^{\rho\rho}\right]^{-1}_{k',k_1}(\mathbf{q}',\mathbf{q}_1)
      \overline{S}_{k_2}(\mathbf{q}_2)
      {\mathfrak M}^{\rho\rho\rho}_{k_1,k_2,k}(\mathbf{q}_1,\mathbf{q}_2,\mathbf{q})\Big], \\
   \label{67}
    G_{k',k}^{(2)}(\mathbf{q}',\mathbf{q};t) & = &
     {\mathfrak M}^{\rho\rho}_{k',k}(\mathbf{q}',\mathbf{q})    \nonumber    \\[2mm]
      & + & \sum_{k_1,\mathbf{q}_1}\sum_{k_2,\mathbf{q}_2}
      \left\langle\rho_{k'}(\mathbf{q}')\right\rangle^t
      \left[{\mathfrak M}^{\rho\rho}\right]^{-1}_{k',k_1}(\mathbf{q}',\mathbf{q}_1)
      {\mathfrak M}^{\rho\rho\rho}_{k_1,k_2,k}(\mathbf{q}_1,\mathbf{q}_2,\mathbf{q}).
 \end{eqnarray}
 Defining $\left[G^{(2)}\right]_{k_2,k}^{-1}(\mathbf{q}_2,\mathbf{q};t)$,
the inverse function of
$G_{k',k}^{(2)}(\mathbf{q}',\mathbf{q};t)$:
 \[
  \sum_{k'',\mathbf{q}''}
  \left[G^{(2)}\right]^{-1}_{k',k''}(\mathbf{q}',\mathbf{q}'';t)
  G^{(2)}_{k'',k_1}(\mathbf{q}'',\mathbf{q}_1;t)=
  \delta_{k',k_1}\delta_{\mathbf{q}',\mathbf{q}_1}\, ,
 \]
one can get from (\ref{65}):
\begin{eqnarray}\label{68}
      \overline{\mu}_{k}(\mathbf{q};t) & = & -\left(\frac{\beta}{S L}\right)^{-1}
      \sum_{k',\mathbf{q}'}
        \left[
          \left\langle\rho_{k'}(\mathbf{q}')\right\rangle^t-\frac{\beta}{S L}G_{k'}^{(1)}(\mathbf{q}')
        \right]\left[G^{(2)}\right]^{-1}_{k',k}(\mathbf{q}',\mathbf{q};t) \nonumber \\
        & + & \frac{3\ii}{2!}\left(\frac{\beta}{S L}\right)^{-1}
            \sum_{k',k''}\sum_{\mathbf{q}',\mathbf{q}''}
             \left\langle\mathbf{p}_{k'}(\mathbf{q'})\right\rangle^t\cdot
              \overline{\mathbf{K}}^{\mathbf{p}\mathbf{p}\rho}_{k',k'',k}(\mathbf{q}',\mathbf{q}'',\mathbf{q};t)\cdot
              \left\langle\mathbf{p}_{k''}(\mathbf{q''})\right\rangle^t,
\end{eqnarray}
where
 \begin{equation}\label{69}
    \overline{\mathbf{K}}^{\mathbf{p}\mathbf{p}\rho}_{k',k'',k}(\mathbf{q}',\mathbf{q}'',\mathbf{q};t)
    =\sum_{k_1,\mathbf{q}_1}{\mathbf{K}}^{\mathbf{p}\mathbf{p}\rho}_{k',k'',k_1}(\mathbf{q}',\mathbf{q}'',\mathbf{q}_1;t)
    \left[G^{(2)}\right]^{-1}_{k_1,k}(\mathbf{q}_1,\mathbf{q};t).
 \end{equation}
Now considering (\ref{68}), (\ref{62}) we can write down the
entropy operator $\hat{S}^{(\mathrm G+1)}(t)$ (\ref{55a}) as follows:
\begin{eqnarray}\label{70}
\lefteqn{   \hat{S}^{(\mathrm G+1)}(t)  = \beta\frac{N}{2S}{\sum_{\mathbf{q}}}'
      \nu(\mathbf{q}|0)+\ln Z_\jell+\beta H }\nonumber \\
        &&\mbox{} +  \sum_{k,\mathbf{q}}\Bigg\{\sum_{k',\mathbf{q}'}
        \left[
          \left\langle\rho_{k'}(\mathbf{q}')\right\rangle^t-\frac{\beta}{S L}G_{k'}^{(1)}(\mathbf{q}')
        \right]\left[G^{(2)}\right]^{-1}_{k',k}(\mathbf{q}',\mathbf{q};t)\rho_{k}(\mathbf{q})
        \nonumber \\
        &&\mbox{} -  \frac{3\ii}{2!}
            \sum_{k',k''}\sum_{\mathbf{q}',\mathbf{q}''}
             \left\langle\mathbf{p}_{k'}(\mathbf{q'})\right\rangle^t\cdot
              \overline{\mathbf{K}}^{\mathbf{p}\mathbf{p}\rho}_{k',k'',k}(\mathbf{q}',\mathbf{q}'',\mathbf{q};t)\cdot
              \left\langle\mathbf{p}_{k''}(\mathbf{q''})\right\rangle^t\rho_{k}(\mathbf{q})\Bigg\}-     \Phi^{(\mathrm G+1)}(t),
  \end{eqnarray}
where $\Phi^{(\mathrm G+1)}(t)$ according to (\ref{55b}) and
(\ref{62}), (\ref{68}) is a function of a higher order with
respect to the parameters of the reduced description
$\left\langle\rho_{k}(\mathbf{q})\right\rangle^t$,
$\left\langle\mathbf{p}_{k}(\mathbf{q})\right\rangle^t$, including
the sixth with respect to
$\left\langle\mathbf{p}_{k}(\mathbf{q})\right\rangle^t$ and its
structure is important in calculating the Kawasaki-Gunton
projection operator~(\ref{6}) in terms of
  $\dfrac{\delta\rho^{(\mathrm G+1)}_{\mathrm{q}}(t)}{\delta \langle \rho_k(\mathbf{q})\rangle^t}$,
  $\dfrac{\delta\rho^{(\mathrm G+1)}_{\mathrm{q}}(t)}{\delta \langle \mathbf{p}_k(\mathbf{q})\rangle^t}$.
Considering the above, we get
\begin{eqnarray}\label{71}
   \ii L_N \rho^{(\mathrm G+1)}_{\mathrm{q}}(t)& = & \sum\limits_{k,\mathbf{q}}
   \int\limits_0^1\dd\tau
   \left[\rho^{(\mathrm G+1)}_{\mathrm{q}}(t)\right]^\tau
   \dot{\rho}_{k}(\mathbf{q})
   \left[\rho^{(\mathrm G+1)}_{\mathrm{q}}(t)\right]^{1-\tau}
   W^{(\mathrm G+1)}_{\rho\rho}(k,\mathbf{q};t)
\nonumber\\
  & + & \sum\limits_{k,\mathbf{q}}
   \int\limits_0^1\dd\tau
   \left[\rho^{(\mathrm G+1)}_{\mathrm{q}}(t)\right]^\tau
   \dot{\mathbf{p}}_{k}(\mathbf{q})
   \left[\rho^{(\mathrm G+1)}_{\mathrm{q}}(t)\right]^{1-\tau}\cdot \mathbf{W}^{(\mathrm G+1)}_{\mathbf{p}\mathbf{p}}(k,\mathbf{q};t),
\end{eqnarray}
where
\begin{eqnarray}\label{73}
   W^{(\mathrm G+1)}_{\rho\rho}(k,\mathbf{q};t) & = & \sum_{k',\mathbf{q}'}
        \left[
          \left\langle\rho_{k'}(\mathbf{q}')\right\rangle^t-\frac{\beta}{S L}G_{k'}^{(1)}(\mathbf{q}')
        \right]\left[G^{(2)}\right]^{-1}_{k',k}(\mathbf{q}',\mathbf{q};t)\nonumber \\
     & - & \frac{3\ii}{2!}
            \sum_{k',k''}\sum_{\mathbf{q}',\mathbf{q}''}
             \left\langle\mathbf{p}_{k'}(\mathbf{q'})\right\rangle^t\cdot
              \overline{\mathbf{K}}^{\mathbf{p}\mathbf{p}\rho}_{k',k'',k}(\mathbf{q}',\mathbf{q}'',\mathbf{q};t)\cdot
              \left\langle\mathbf{p}_{k''}(\mathbf{q''})\right\rangle^t,\\
   \label{74}
   \mathbf{W}^{(\mathrm G+1)}_{\mathbf{p}\mathbf{p}}(k,\mathbf{q};t) & = &
   \sum_{k',\mathbf{q}'}
          \left\langle\mathbf{p}_{k'}(\mathbf{q}')\right\rangle^t\cdot
        \left[\mathbf{G}^{(3)}\right]^{-1}_{k',k}(\mathbf{q}',\mathbf{q};t).
\end{eqnarray}

\ukrainianpart

\title{В'язко-еластичний опис нерівноважної електронної підсистеми напівобмеженого металу \\ в узагальненій моделі ``желе''}
\author{П. П. Костробій\refaddr{label1}, Б. М. Маркович\refaddr{label1}, А. І. Василенко\refaddr{label2}, М. В. Токарчук\refaddr{label1,label2}}
\addresses{
\addr{label1} Національний університет ``Львівська політехніка'', вул. С. Бандери, 79013 Львів, Україна
\addr{label2} Інститут фізики конденсованих систем НАН України, вул. I. Свєнціцького, 1, 79011 Львів, Україна
}
%
%
\makeukrtitle

\begin{abstract}
\tolerance=3000%
Запропоновано в'язко-еластичний опис електронної  підсистеми напівобмеженого металу на основі  узагальненої моделі
``желе'' із застосуванням методу нерівноважного статистичного
оператора Зубарєва. Отримано нерівноважний статистичний оператор та відповідні узагальнені рівняння переносу для нерівноважних
середніх значень операторів густин числа електронів та їх імпульсу у гауcовому та вищому за ним  наближені, що відповідає кумулянтним середнім третього порядку при розрахунку квазірівноважної статистичної суми
методом функціонального інтегрування.
\keywords узагальнена модель ``желе'', нерівноважний статистичний оператор Зубарєва,
напівобмежений метал, рівняння переносу, квазірівноважна статистична сума
\end{abstract}


\begin{thebibliography}{99}
\bibitem{l1} Nanotechnology Research Directions, IWGN Workshop Report,
Vision for Nanotechnology Research in the Next Decade. Eds.
M.C.~Roco, S.~Williams, P.~Alivisatos. Kluwer Academic Publ.,
2000.


 \bibitem{l11} Kiselev~V., Krylov~O.,
          Adsorption and catalysis on transition metals and their
          oxides. Springer-Verlag, Berlin, 1989. 


\bibitem{l13} Slinko~M.M., Jaeger~N.I. eds.,
Oscillating Heterogeneous Catalytic Systems. In: Studies in Surface Science and Catalysis, Vol.~86. Elsevier, Amsterdam, 1994.

 \bibitem{l14} Naumovets~A.G., Zhang~Zh.,
           Surf. Sci., 2002, \textbf{500}, No.~1--3, 414;
           \doi{10.1016/S0039-6028(01)01539-4}. 

 \bibitem{l15} Tsong~T.T.,
          Prog. Surf. Sci.,  2001, \textbf{67}, 235;
          \doi{10.1016/S0079-6816(01)00026-0}. 

 \bibitem{l16} Gomer~R.,
 Rep. Prog. Phys., 1990, \textbf{53}, 917; \doi{10.1088/0034-4885/53/7/002}. 


 \bibitem{l17} Stoltze~P.,
         Prog. Surf.  Sci., 2000, \textbf{65}, 65; \doi{10.1016/S0079-6816(00)00019-8}. 

\bibitem{Kost000} Kostrobij~P.P., Markovych~B.M., Tokarchuk~M.V., Ignatyuk~V.V., Hnativ~B.V.,
Reaction-diffusion processes in systems ``metal--gas''.
Lviv Polytechnic National University, Lviv, 2009 (in Ukrainian).


\bibitem{Ign000} Ignatyuk~V.V.,
Phys. Rev. E, 2009, \textbf{80}, No.~4, 041133(13);
\doi{10.1103/PhysRevE.80.041133}.

 \bibitem{TDF0} Runge~E., Gross~E.K.U.,
 Phys. Rev. Lett., 1984, \textbf{52}, 997; \doi{10.1103/PhysRevLett.52.997}.

\bibitem{TDF1} Ullrich~C.A., Grossmann~U.I., Gross~E.K.U.,
 Phys. Rev. Lett., 1995, \textbf{74}, No.~6, 872; \\ \doi{10.1103/PhysRevLett.74.872}. 

\bibitem{TDF2} Vignale~G., Konh~W.,
 Phys. Rev. Lett., 1996, \textbf{77}, No.~10, 2037; \doi{10.1103/PhysRevLett.77.2037}. 

\bibitem{TDF3} Vignale~G., Ullrich~C.A.,
 Phys. Rev. Lett., 1997, \textbf{79}, No.~24, 4878;  \doi{10.1103/PhysRevLett.79.4878}. 


\bibitem{TDF6} Maitra~N.T., Burke~K., Appel~H., Gross~E.K.U., Van Leeuwen~R.,
Ten topical questions in time-dependent density functional theory.
In: Parr R.G., Sen K.D. eds., Reviews of modern quantum chemistry:
a celebration of the contributions of Robert G. Parr.
World-Scientific, 2001.

\bibitem{TDF7} Ullrich~C.A., Vignale~G.,
Phys. Rev. B, 2002, {\bf 65}, 245102; \doi{10.1103/PhysRevB.65.245102}.

\bibitem{TDF8} Maitra~N.T., Burke~K., Woodward~C.,
Phys. Rev. Lett., 2002, \textbf{89}, No.~2, 023002; \\ \doi{10.1103/PhysRevLett.89.023002}.

\bibitem{TDF11} Onida~G., Reining~L., Rubio~A.,
Rev. Mod. Phys.,  2002, \textbf{74}, No.~2, 601; \\ \doi{10.1103/RevModPhys.74.601}. 

\bibitem{TDF12} Tokatly~I.V., Pankratov~O.,
Phys. Rev. B, 2003, {\bf 67}, 201103(R); \doi{10.1103/PhysRevB.67.201103}.

\bibitem{TDF14} Marques~M.A.L., Gross~K.M.,
 Annu. Rev. Phys. Chem., 2004, \textbf{55}, 427;
\\ \doi{10.1146/annurev.physchem.55.091602.094449}. 

\bibitem{TDF15}  Botti S., Sottile F., Vast~N., Olevano~V., Reining~L.,
Weisske~H.-C., Rubio~A., Onida~G., Del Sole~R., Godby~R.W.,
Phys. Rev. B, 2004, \textbf{69}, No.~15, 155112; \doi{10.1103/PhysRevB.69.155112}. 

\bibitem{TDF18} Dion M., Burke K.,
Phys. Rev. A, 2005, \textbf{72}, 020502; \doi{10.1103/PhysRevA.72.020502}. 

\bibitem{TDF20}  Burke~K., Werschnik J., Gross~E.K.U.,
 J. Chem. Phys., 2005,  {\bf 123}, 062206; \doi{10.1063/1.1904586}.

\bibitem{EQ1} Equiluz A., Xing~S.C., Quinn~J.J.,
Phys. Rev. B, 1975, \textbf{11}, No.~6, 2118;  \doi{10.1103/PhysRevB.11.2118}. 

\bibitem{EQ2} Equiluz A., Quinn~J.J.,
Phys. Rev. B, 1978, \textbf{14}, No.~4, 1347; \doi{10.1103/PhysRevB.14.1347}. 

\bibitem{EQ3} Equiluz A.,
 Phys. Rev. B, 1979, \textbf{19}, No.~4, 1689; \doi{10.1103/PhysRevB.19.1689}. 

\bibitem{Grif} Griffin Al., Zaremba E.,
Phys. Rev. A, 1973, \textbf{8}, No.~1, 486; \doi{10.1103/PhysRevA.8.486}. 


\bibitem{Kost5} Kostrobii P.P., Markovych B.M., Rudavskii Yu.K., Tokarchuk M.V.,
Condens. Matter Phys., 2001, \textbf{4}, No.~3(27), 407. 

\bibitem{Kost111} Kostrobii P.P., Markovych B.M., Vasylenko A.I., Tokarchuk M.V.,
 Ukr. J. Phys., 2007, \textbf{52}, No.~11,~1096. 


\bibitem{l121} Zubarev D.N., Itogi Nauki i Tekhniki. Sovremennye Problemy Matematiki, 1980, \textbf{15} 131, 
(in Russian).

 \bibitem{l122}Zubarev D., Morozov V., R\"opke G., Statistical Mechanics of
 Nonequilibrium Processes. Akademie Verlag, Berlin, 1996. 


\bibitem{l184}  Kostrobij~P.P., Markovych~B.M.,
J.~Phys. Stud., 2003, \textbf{7},   No.~2, 195, (in Ukrainian). 

\bibitem{l185} Kostrobij~P.P., Markovych~B.M.,
J.~Phys. Stud., 2003, \textbf{7},   No.~3, 298, (in Ukrainian). 


\bibitem{sss1} Lang~N.D., Kohn~W.,
Phys. Rev. B, 1971, \textbf{3}, No.~4, 1215; doi{10.1103/PhysRevB.3.1215}. 


\bibitem{sss2} Fiolhais~C., Henriques~C., Sarr\'{i}a~I, Pitarke~J.M.,
 Prog. Surf. Sci., 2001, \textbf{67}, 285; \\ \doi{10.1016/S0079-6816(01)00030-2}.

\bibitem{sss3} Rose~J.H., Dobson~J.F.,
Solid State Commun., 1981, \textbf{37}, No.~2, 91; \doi{10.1016/0038-1098(81)90719-5}. 

 \bibitem{KMVT2011} Kostrobij P.P., Markovych B.M., Vasylenko A.I., Tokarchuk M.V.,
Ukr. J.~Phys., 2011, \textbf{56}, No.~2, 179. 

\bibitem{Kost22}Kostrobij~P.P., Markovych~B.M.,
Condens. Matter Phys., 2003, \textbf{6}, No.~2(34), 347. 

\end{thebibliography}
\end{document}